\documentclass[epjST]{svjour}
\usepackage{graphics}
\usepackage{color}
\usepackage{amsmath}

\definecolor{pdcolor}{rgb}{1,0.5,0}

\definecolor{pdblue}{rgb}{0,0,1}

\definecolor{rkgreen}{rgb}{0,1,0}

\begin{document}
\title{Energy-dependent diffusion in a soft periodic Lorentz gas}
\author{S. Gil-Gallegos\inst{1}\fnmsep\thanks{\email{s.gilgallegos@qmul.ac.uk}}
         \and R. Klages\inst{1,2,3}\fnmsep\thanks{\email{r.klages@qmul.ac.uk}}
         \and J. Solanp{\"a}{\"a}\inst{4}\fnmsep\thanks{\email{janne@solanpaa.fi}}
         \and E. R{\"a}s{\"a}nen\inst{4}\fnmsep\thanks{\email{esa.rasanen@tut.fi}} }
\institute{Queen Mary University of London, School of Mathematical Sciences, Mile End Road, London E1 4NS, UK
      \and Institut f\"{u}r Theoretische Physik, Technische Universit\"{a}t Berlin, Hardenbergstra{\ss}e 36, 10623 Berlin, Germany
      \and Institute for Theoretical Physics, University of Cologne, Z\"ulpicher Stra{\ss}e 77, 50937 Cologne, Germany
      \and Laboratory of Physics, Tampere University of Technology, Tampere FI-33101, Finland}

    \abstract{The periodic Lorentz gas is a paradigmatic model to
        examine how macroscopic transport emerges from microscopic
        chaos. It consists of a triangular lattice of circular hard
        scatterers with a moving point particle. Recently this system
        became relevant as a model for electronic transport in
        low-dimensional nanosystems such as molecular graphene.
      However, to more realistically mimic such dynamics, the hard
      Lorentz gas scatterers should be replaced by soft potentials.
      Here we study diffusion in a soft Lorentz gas with Fermi
      potentials under variation of the total energy of the moving
      particle. Our goal is to understand the diffusion coefficient as
      a function of the energy. In our numerical simulations we
      identify three different dynamical regimes: (i) the onset of
      diffusion at small energies; (ii) a transition where for the
      first time a particle reaches the top of the potential,
      characterized by the diffusion coefficient abruptly dropping to
      zero; and (iii) diffusion at high energies, where the diffusion
      coefficient increases according to a power law in the energy.
      All these different regimes are understood analytically in terms
      of simple random walk approximations.  } 
\maketitle
\section{Introduction}
\label{intro}
The need for smaller and more efficient electronic devices challenges
both science and technology by pushing their boundaries.  In order to
understand the structure and dynamics of matter on very small scales,
simple mathematical models have been constructed.  An important
example concerns the transport of matter by diffusion: Typically this
problem is studied by methods of conventional non-equilibrium
statistical mechanics, where a stochastic process is assumed to govern
the collisions between particles.  However, starting from microscopic
deterministic equations of motion helps to understand the origin of
diffusion within the framework of dynamical systems theory.
Employing this approach it is possible to derive a macroscopic
irreversible process out of reversible microscopic equations of motion
\cite{Do99,Gas98,Kla07}.

By understanding diffusion in simple dynamical systems it is possible
to learn about the transport properties of more complex models. For
example, in piecewise linear one-dimensional chaotic maps it has been
found that the diffusion coefficient is a fractal function of control
parameters \cite{KlaDor,KlDo99,GrKl02}. A relatively simple
two-dimensional map with similarly irregular diffusion coefficients is
the standard map~\cite{Chir79,RRW81}, which can be derived as
a time discrete version of a chaotic nonlinear pendulum equation
\cite{BlGJ96,CGMR15}.  Here it has been observed that normal diffusion
is interrupted by regions in parameter space which are related to
accelerator modes yielding superdiffusive
transport~\cite{ZGNR86,MaRo14,Zas02}.  Normal diffusion means that the
mean square displacement of an ensemble of particles increases
linearly with time while superdiffusion refers to a time dependence
that grows faster than linear in time. A nonlinear time dependence of
the mean square displacement is generally called anomalous diffusion
\cite{KRS08}.

Another class of dynamical systems with highly interesting transport
properties is given by Hamiltonian particle billiards \cite{Sza00}.
In the paradigmatic two-dimensional Lorentz gas a point particle
performs specular reflections with circular scatterers distributed in
the plane~\cite{Lo05}.  Originally constructed to model the motion of
electrons in metals, the Lorentz gas has been widely investigated from
both mathematical and physical points of
view~\cite{Do99,Gas98,Kla07,BS81,GN90}.  In periodic versions of the
Lorenz gas it has been found that the diffusion coefficient is an
irregular function of the density of scatteres as a parameter
\cite{KlDe00,HaKlGa02,KlKo02}.  This relates to the line of work on
irregular parameter-dependent diffusion coefficients in simple maps
referred to above. However, it is an open question whether the
diffusion coefficient in the periodic Lorentz gas exhibits a fractal
structure under variation of control parameters \cite{Kla07,Det14}.  A
related billiard is a one-dimensional corrugated floor, where a
particle experiences a vertical force and collides with arcs
positioned horizontally \cite{HaGa01}. Here the diffusion coefficient
was numerically computed as a function of the energy, and islands in
phase space were found for particular energies yielding
superdiffusion. When no islands were present diffusion was normal.
Altogether the diffusion coefficient displayed an intricate dependence
on the energy.  A related systems was the bouncing ball billiard,
where a particle diffused on a vibrating corrugated floor by losing
energy at collisions. Again diffusion was detected numerically to be
highly irregular under variation of the particle's energy
\cite{MaKl03}.  All these studies point to the conjecture that under
certain conditions the diffusion coefficient in deterministic
dynamical systems may exhibit a non-trivial, often fractal-like
dependence on control parameters~\cite{KlDo99,Kla07,GaKl}.

To model more realistic particle collisions, the hard Lorentz gas
scatterers should be softened. When the walls of a scattering
  billiard are smoothened by using a soft repulsive potential,
  periodic orbits appear near a special class of trajectories in the
  original system~\cite{TurRom98,RomTur99}. These new periodic orbits
  are accompanied by periodic islands of stability yielding a mixed
  phase space.  As a mixed phase space is decomposable into disjoint
  invariant sets, ergodicity is no longer present in such systems
  \cite{LiLi92}. Singular trajectories inducing islands of stability
in the phase space have also been reported for atom-optic billiards
studied both numerically and experimentally \cite{KFAD04}.  For
  periodic billiards with scatterers softened by attractive Coulombic
  potentials, a mathematical proof states that for energies above a
  certain threshold the motion of particles is
  diffusive~\cite{Kna87}. By exploring the dependence of the
diffusion coefficient $D$ on the energy $E$ in such a system a
relation of $D(E) \sim E^{3/2}$ for large energies has been derived
\cite{Nob95}. Moreover, a soft inelastic periodic Lorentz gas with
time-dependent scatterers was proposed in
Ref.~\cite{aguer2010elastic}. Here it was found that the diffusion
coefficient grows like $D(E)\sim E^{5/2}$ for large energies. In
  both cases results from computer simulations were supported by
  constructing different simple random walk models from which the
  diffusion coefficient could be calculated analytically as a function
  of the energy.

Different models were considered to replace the hard walls in Lorentz
gases by soft potentials, most notably by using trigonometric
functions yielding an egg-crate potential
\cite{BZM85,GZR87,GZR88,GWNO90,Pan}, by a Lennard-Jones potential
\cite{YaZh}, or simply by a finite potential height \cite{Bal88}.  The
first type of systems was designed to model experiments on electrons
in lateral superlattices under magnetic fields \cite{GZR88}.
  Here it was experimentally observed that the magnetoresistance
  varies highly irregularly as a function of the magnetic field
  strength \cite{LKP91,Weis91}. It was then shown in computer
  simulations combined with dynamical systems and stochastic theory
  that specific peaks in the magnetoresistance are due to periodic
  orbits with associated islands of stability in phase space, which
  correspond to electrons circling around specific sets of
  scatterers \cite{FGK92,FlSS96}. A similar type of system that
attracted much attention recently is artificial graphene, where
electrons are confined to a hexagonal configuration of scatterers by
using, e.g., semiconductor quantum dots~\cite{Gibertini,Rasanen} or
molecules on a metallic surface~\cite{GMKGM12,PRNAR16}. The latter
system is often referred to as molecular graphene, which is
topologically equivalent to the triangular Lorentz gas. Hence, the
study of periodic billiard systems with soft potentials has direct
relevance not only in the theory of dynamical systems and diffusion,
but also in present nanotechnology.

In this work we study the diffusion coefficient as a function of the
energy as a control parameter in a soft Lorentz gas. While in a
conventional hard-wall system the energy generates only a trivial
scaling of the diffusion coefficient, as the velocity is constant, the
scenario is completely different for a soft potential.  In a recent
work by the present authors~\cite{KGSSR18} it was shown that a soft
Lorentz gas modeled by Fermi potentials exhibits an intricate
interplay between normal and anomalous diffusion as a function of the
density of scatterers. Here we investigate the same system,
  however, we are now interested in its diffusive properties under
  variation of the energy of the moving particle as a control
  parameter while we keep the density of scatterers fixed.  We resort
to extensive numerical simulations supplemented by analytical random
walk approximations to explain our numerical results.

Our article is organized as follows: In Sec.~\ref{sec:model} we
construct a softened version of the Lorentz gas and describe our
numerical methods.  In Sec.~\ref{sec:randomwalksenergy} we present our
simulation results in the small-energy regime, where we develop {\it
  ad hoc} random walk approximations for the diffusion coefficient
based on a simple phase space argument~\cite{MaZw83} as well as on a
collision length approach.  In Sec.\ref{sec:transition} we explore an
intermediate regime where the energy is close to the maximum of a
scatterers and derive another random walk model.  In
Sec.~\ref{HERnumerics} we characterize the asymptotic form of the
diffusion coefficient for high energies. We also discuss the
complicated structure of the phase space and the appearance of islands
of stability by varying the energy parameter.  Finally, in
Sec.~\ref{sec:conclusions} we conclude our work and present open
questions.

\section{Model and numerical methods}
\label{sec:model}

\subsection{Soft Lorentz gas}

\begin{figure}
\centering
\begin{minipage}[c]{0.49\textwidth}
\centering 
\resizebox{\columnwidth}{!}{ \includegraphics{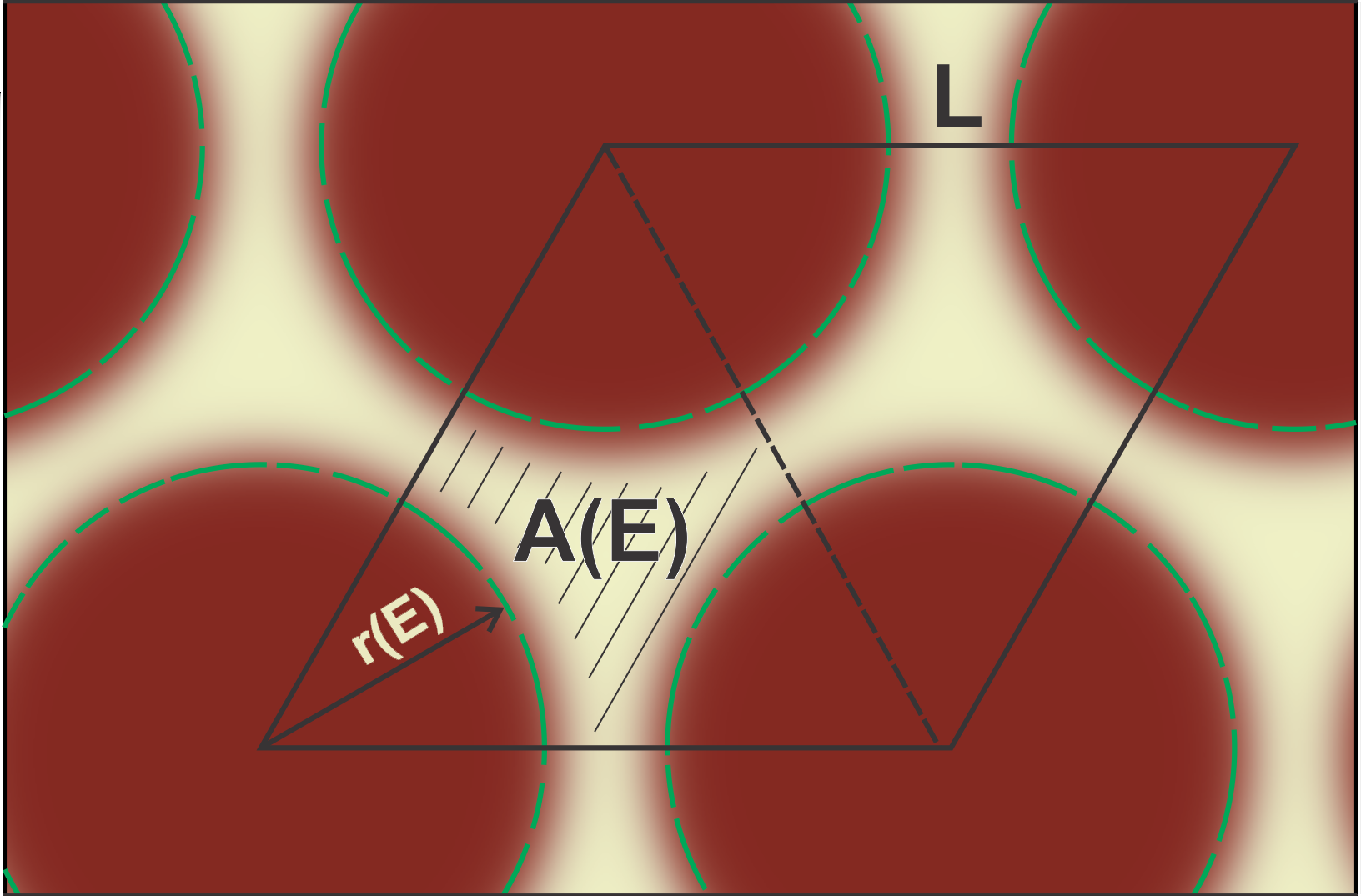} }
\end{minipage}
\begin{minipage}[c]{0.5\textwidth}
\resizebox{\columnwidth}{!}{ \includegraphics{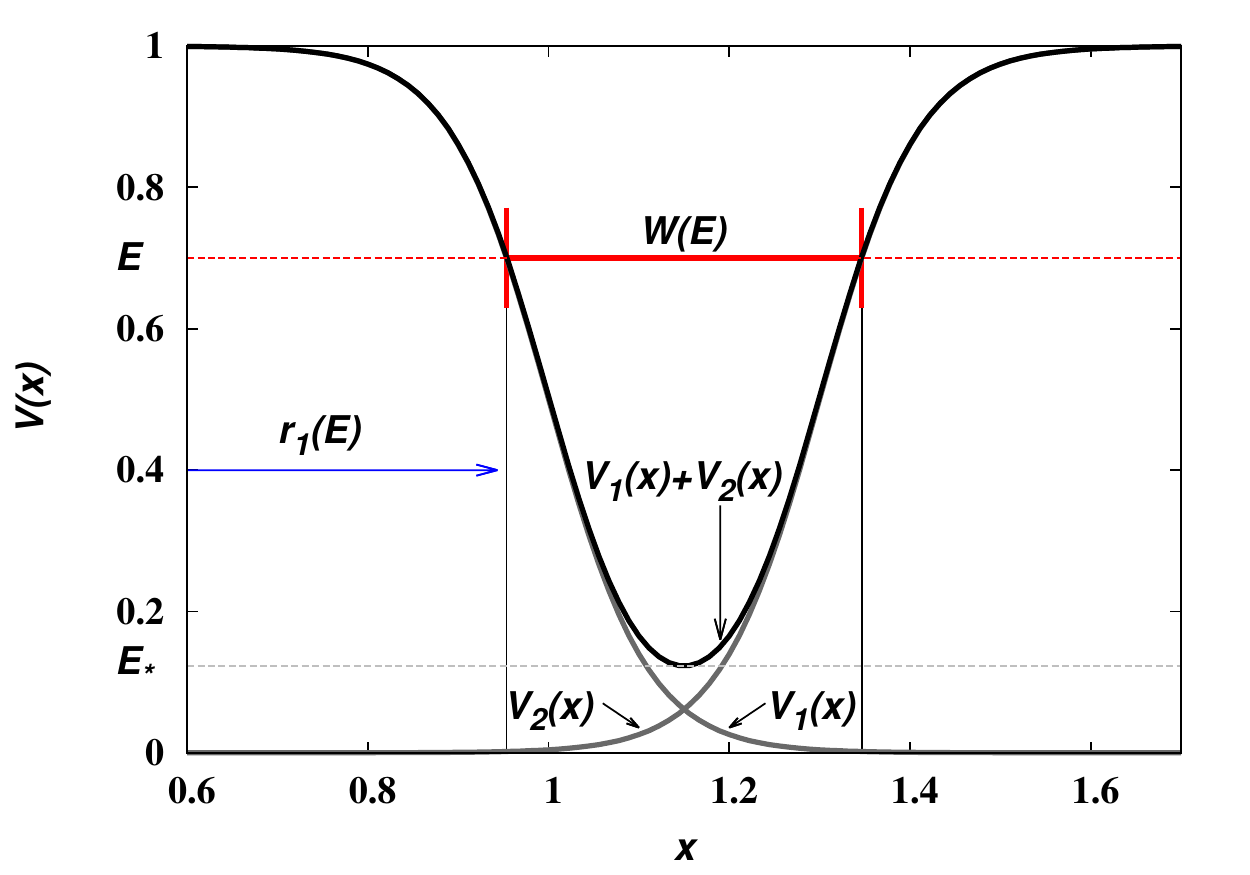} } 
\end{minipage}
\caption{Left panel: Profile of the potential of our soft Lorentz
  gas defined in Eqs.~\eqref{fermi_pot} and~\eqref{SumFermi_pot}.
  $A(E)$ is the area inside a unit cell enclosed by equipotential
  lines with radius $r(E)$, where $E$ is the total energy of a
  particle.  Right panel: Potential $V=V_1+V_2$ along the $x$-axis
  (bottom boundary of the parallellogram unit cell in the left panel).
  The gap size $W(E)$ and the radius $r_1(E)$ are defined by the
  energy $E$, cf.\ Eqs.~(\ref{WEi}) and~(\ref{radiusE}). $E_*$ holds
  for the energy threshold as explained in the text.  }
\label{w_V1}
\end{figure}

For our study we use the soft model introduced in Ref.~\cite{KGSSR18},
which has the advantage that it reproduces the conventional triangular
Lorentz gas with hard circular scatterers in a specific limit of the
softness parameter. For this model each circular scatterer is defined
by a Fermi potential of the form
\begin{equation}
V_{\rm F}(r)=\frac{1}{1+\exp\big(\frac{|\mathbf{r}|-r_o}{\sigma}\big)}, 
\label{fermi_pot}
\end{equation}
where $\sigma$ and $r_o$ are the softness and the radius of a
scatterer, respectively. The complete potential field consisting of an
infinite triangular array of scatterers is given by
\begin{equation}
V(\mathbf{r})=\sum \limits_{n}  V_{\rm F}(\mathbf{r}-\mathbf{r}_n), 
\label{SumFermi_pot}
\end{equation}
where $\mathbf{r}_n$ is the position vector to the $n$-th point of the
lattice in the plane.  The maxima and minima of the potential are
located at the vertices and centers of the triangles, respectively,
whereas the edges of the triangles have saddle points; see
Fig.~\ref{w_V1}.

If the particle has total energy $E>E_*$ it can escape from the region
enclosed by three scatterers through the available space between
$r(E)$ and $r(E)+W(E)$, where the gap $W(E)$ is given by
\begin{equation}
W(E)=L-2r(E).
\label{WEi}
\end{equation}
Here $L$ is the fixed distance between the centres of two adjacent scatterers. 
This scenario is visualized in Fig.~\ref{w_V1}. 

The exact radius $r$ should be computed by considering the
contribution of all potentials located at each point of the lattice.
If we use the contribution of two adjacent potentials as shown in
Fig.~\ref{w_V1}, it would be necessary to solve for $r(x)$ and
$r(x-L)$ when $V_1(x)+V_2(x)=E$.  Alternatively, considering only
$V_1(x)$ we can obtain an analytical expression for the radius.  By
solving $V_1(x)=E$ for $r(x)$ we get the radius in terms of the
energy.  For simplicity we will label this new radius with $r_1$, and
it is given by
\begin{equation} 
r_1(E)=\sigma \ln(1/E-1)+r_o.
\label{radiusE} 
\end{equation}
Note that due to the overlapping potentials the radius $r_1$ is an
underestimation of the real radius $r$ in Eq.~(\ref{WEi}), since $r_1$
has been computed considering only one potential. Hence, the real gap
size $W(E)$ given by Eq.~(\ref{WEi}) with $r_1$ yields an
overestimation. This effect is enhanced when $E\to E_*$, as the
difference $r-r_1$ increases.

Finally, we connect $W(E)$ with the separation of the scatterers in
the hard Lorentz gas $w$.  These two quantities are related by the
lattice length
$$L=2r_o+w=2r(E)+W(E),$$ 
where $r_0$ is the radius of a hard Lorentz gas scatterer and $w$ is
the minimal distance, or gap size, between two adjacent scatterers.
Therefore we can use either the parameter $L$ or $w$ to define the
density of the scatterers.  In our random walk model introduced in
Sec.~\ref{sec:randomwalksenergy} we use the radius of
Eq.~(\ref{radiusE}), bearing in mind that there is a greater error the
closer $E$ is to the threshold $E_*$. In this work we set the softness
parameter in Eq.~(\ref{fermi_pot}) to $\sigma=0.01$ and the lattice
distance to $L=2r_o+w$ with $r_o=1$ and $w=0.05$. These values enable
normal diffusion in a relatively large interval of energies as
discussed in detail in Sec.~\ref{HERnumerics}.

\subsection{Computation of the diffusion coefficient}

The diffusion coefficient is defined by \cite{Do99,Gas98,Kla07}
\begin{equation}
D=\lim_{t \rightarrow \infty} \frac{\langle(\textbf{r}(t)-\textbf{r}(0))^2\rangle}{4t},
 \label{diffusion}
\end{equation}  
where $\textbf{r}(t)$ is the position of a particle at time $t$ and
$\langle(\textbf{r}(t)-\textbf{r}(0))^2\rangle$ yields the mean square
displacement (MSD) with the ensemble average of particles denoted by
the angular brackets. If the limit in Eq.~(\ref{diffusion}) exists
or, equivalently, the growth of the MSD is linear in time, we have
normal diffusion.

According to Eq.~(\ref{diffusion}) one can obtain $D$ from
  calculating the MSD. For our model we first computed the MSD from
  computer simulations, which were carried out with the bill2d
  software package \cite{solanpaa2016bill2d} at different values of
  the energy. The force acting on the particles is given by the
gradient of Eq.~(\ref{SumFermi_pot}). Here we sum only over lattice
points of scatterers that are in a rhomboid unit cell, see
Fig.~\ref{w_V1}.  An ensemble of particles is uniformly distributed in
the coordinate space associated to such a unit cell. Trivially,
  only energetically valid combinations of $(x,y,v_x,v_y)$ are
  allowed.  In an ergodic dynamical system the selection of initial
  conditions is not important, as in the long run all regions in phase
  space will be sampled for any given initial condition \cite{Do99}.
  However, since in Hamiltonian dynamical systems with a mixed phase
  space ergodicity is broken, cf.\ our discussion in Sec.~\ref{intro},
  we need to make sure that our ensemble of initial points is large
  enough such that it adequately samples different disjoint regions in
  phase space to reflect the full dynamics \cite{LiLi92}.

\begin{figure}
	\begin{center}		
\rotatebox{-90}{\resizebox{0.5\columnwidth}{!}%
{\includegraphics{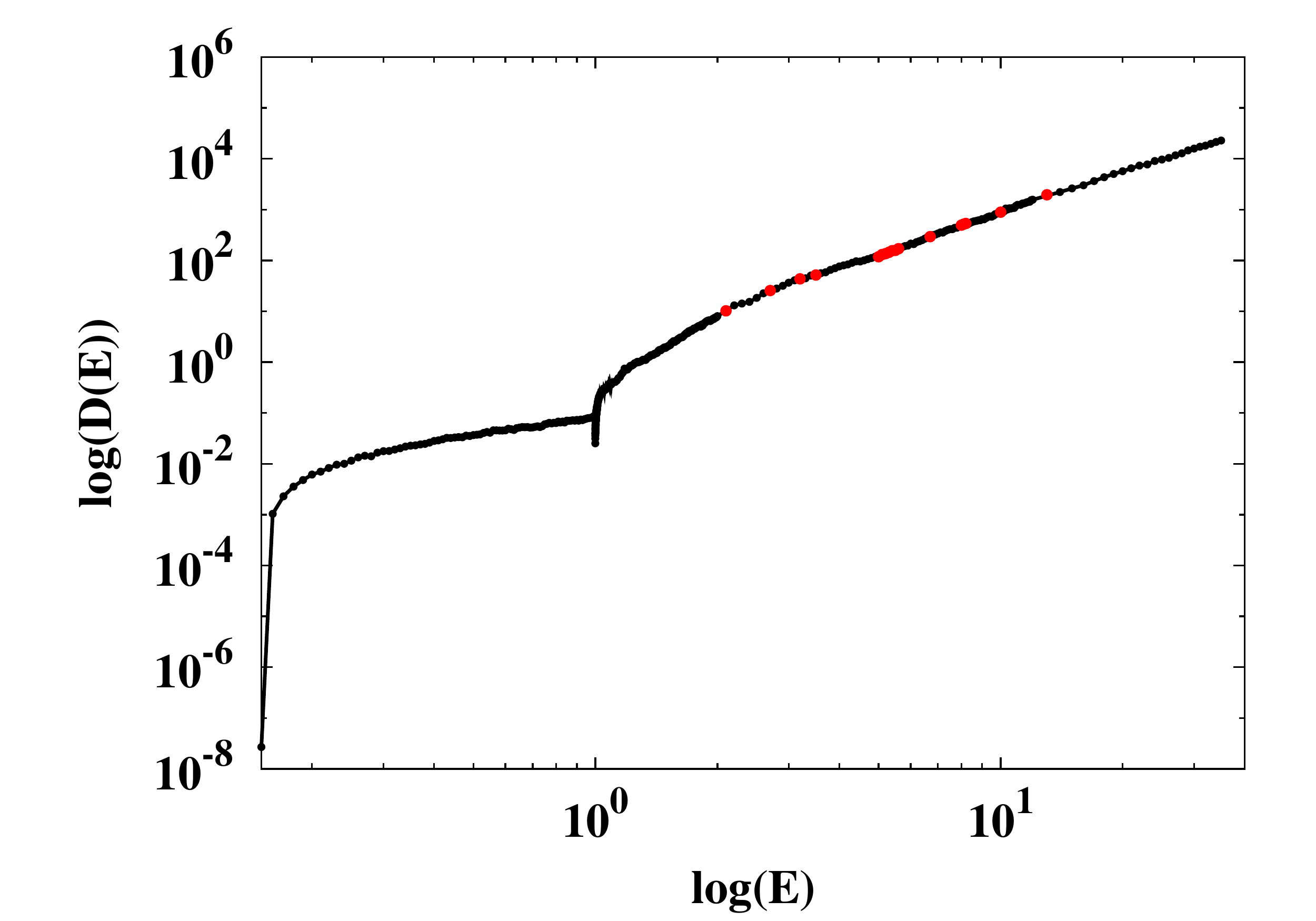}}}
	\end{center}        
	\caption{Diffusion coefficient $D$ as a function of energy $E$
          in a double-logarithmic plot obtained from simulations
          according to Eq.~(\ref{diffusion}); see the text for the
          simulation parameters. Red circles denote energies where the
          model exhibits superdiffusion, i.e., $D(E)$ does not exist.}
	\label{all}
\end{figure}

For reliably extracting the MSD from simulations we use the following
input parameters: For energies $E<2$ the computing time is $t=5000$
and the ensemble size of the particles $N=20000$. For energies $E>2$
we set the ensemble size to $N=40000$ and the iteration time to
$t=40000$. The time step is $\Delta t 10^{-3}$. By increasing
the ensemble size, using a smaller time step, etc., we have tested
that these parameter values yield good convergence of the MSD to an
asymptotic limit where it grows linearly in time $t$ in parameter
regions $(w,\sigma)$ where islands in phase space do not seem to exist
\cite{thesisGil}.
In Fig.~\ref{all} we present results for the diffusion coefficient
obtained from simulations.  We can distinguish two different regimes
for small and large energies, as well as a distinctive transition
point around $E=V_{\rm max}=1$. In the following we will analyse these
different regimes in full detail. We also note that there are
  parameter values where the diffusion coefficient was not computable
  from simulations, that is, the MSD did not reach a linear regime
  within the computing time. These parameter regions where the MSD
  grows faster than linear are marked as red dots.  We elaborate on
  the physical mechanism generating this specific dynamics in
  Sec.~\ref{HERnumerics} below.

\section{Diffusion coefficient for small energies}
\label{sec:randomwalksenergy}

In previous literature random walks have been used as simple
  models to reproduce analytically computer simulations results for
  the diffusion coefficient in the hard Lorentz gas as a function of
  the density of scatterers \cite{MaZw83,KlDe00,KlKo02}. This
  comparison clarifies whether the deterministic diffusive dynamics
  generated in the chaotic Lorentz gas can be understood in terms of a
  simple stochastic process, which elucidates the origin and the type
  of the diffusive dynamics. On this basis similarities and
  differences between deterministic and stochastic diffusive processes
  can be explored \cite{Do99,Gas98,Kla07}. In this section we test to
  which extent the diffusion coefficient in our soft Lorentz gas can
  be explained by simple random walks for small energies, i.e., close
  to the onset of diffusion. For this purpose we construct a random
  walk model for diffusion as a function of the energy and compare our
  analytical results to computer simulations.

\subsection{Analytical random walk approximations}
For small energies $E<V_{\rm max}=1$ we first approximate the
diffusion coefficient as a function of the total energy by means of a
phase space argument~\cite{MaZw83,KlDe00,KlKo02}, and secondly
employing a collisionless flight or Boltzmann approach \cite{KlDe00};
see also Ref.~\cite{KGSSR18}.  For random walks on two-dimensional
lattices the diffusion coefficient is obtained from
\begin{equation}
D=\frac{l^2}{4\tau},
\label{Dif}
\end{equation}
where $l$ is the distance travelled in one step during the time
interval $\tau$ of the random walk. 

In order to construct a random walk model by using this equation,
  the starting point is to identify what we call a ``trap'' in our
  model. In the soft Lorentz gas this is the region that is available
  for a particle in the position space inside a triangular unit cell,
  orresponding to what was called a trap in the hard Lorentz
  gas~\cite{MaZw83,KlDe00,KlKo02}. As is shown in the left panel of
  Fig.~\ref{w_V1} there are three exits from a trap.  The width of an
  exit, respectively the minimal gap size between two nearby
  scatterers, is defined in the right panel of Fig.~\ref{w_V1}.

\subsubsection{Machta-Zwanzig random walk and phase space argument}

According to a random walk approximation put forward by Machta and
Zwanzig \cite{MaZw83}, a particle travels from one trap to another by a
random walk yielding the diffusion coefficient of Eq.~(\ref{Dif}).
For a soft Lorentz gas the escape time $\tau$ expressed in
terms of the energy $E$ leads to
\begin{equation}
D_{\rm MZ}(E)=\frac{l^2}{4\tau(E)}, 
\label{D_MZ}
\end{equation}
where the distance $l$ between two adjacent traps is constant, as we
are not including a variation of the lattice. Hence, $l$ can be
calculated by geometrical means.  The average escape time from a trap
$\tau^{-1}$ is given by the quotient of the total phase space $\Omega$
and the fraction of phase space $\omega$ that escapes from the gap
during time $\tau$,
\begin{equation}
\tau^{-1}=\frac\omega  \Omega.
\label{tauE}
\end{equation}
Here the velocity space is $2\pi v(E)$, where $v(E)$ is to be
determined. The total volume of the phase space is thus
\begin{equation}
\Omega= A_{\rm trap}2\pi v(E),
\label{OmegaE}
\end{equation}
where $A_{\rm trap}$ is the area available in position space which is
also a function of the energy, $A_{\rm trap}=A(W(E)).$ This area
depicted in Fig.~\ref{w_V1} depends on the radius $r_1(E)$ as
expressed by Eq.~(\ref{radiusE}).

We can calculate the available area for a particle in the position
space by geometrical means in analogy to the conventional (hard-wall)
Lorentz gas. The area is determined by the equipotential lines with
$V_1(r)=E$ and the three exits of the trap.  We take the area of the
unit cell and subtract three times the area formed by semicircles with
radius $r_1(E)$ as in Eq.~(\ref{radiusE}). We find that $A(W(E))$,
simplified to $A(E)$, is given by the expression
$$A(E)=\sqrt3(2r_o+w/2)^2-\frac{\pi}{2}(\sigma \ln (1/E-1)+r_o)^2.$$
Strictly speaking $A(E)$ has a more complicated dependence on $E$ due
to the overlapping of the potentials. However, here we use $A(E)$ as a
first approximation.
where  is the angle between the velocity vector at the moment of exiting the unit cell and the
normal to the boundary dening the gap; the speed jvj = v needs to be determined.
Let us suppose that the velocity $v$ of a particle is constant at the moment of
exiting a trap. Then the particle flux is given by
\begin{equation}
\int |\textbf{n}||\textbf{v}|  \cos  \theta v d \theta = 2v^2,
\end{equation}
where $\theta$ is the angle between the velocity vector at the moment
of exiting the unit cell and the normal to the boundary defining the
gap; the speed $|\textbf{v}|=v$ needs to be determined.  There are
three exits of width $W(E)$ leading to
\begin{equation}
\omega= 3 W(E) 2 v^2(E).
\label{omegaMZ}
\end{equation}
Substituting these quantities into Eq.~(\ref{tauE}) and proceeding
with Eq.~(\ref{D_MZ}) yields our final result
\begin{equation}
D_{\rm MZ}(E)=\frac{3l^2W(E)}{4\pi A(W(E))}v(E).
\label{D_rw}
\end{equation}

\subsubsection{Random walk approximation based on collisionless flights}

We now construct a second random walk approximation, which is based on
collisionless flights. The starting point is the formula
\begin{equation}
D_{\rm B}=\frac{l^2_c}{4\tau_c}
\label{D_b}
\end{equation}
for the diffusion coefficient, where $\tau_c$ stands for the mean free
time between collisions and $l_c$ is the corresponding mean free path
between collisions. Note that in systems composed of smooth
  potentials strictly speaking collisions are not defined, as at any
  instant of time there is a force acting on the particle.  For the
  following approximate analytical calculations, we thus define a
  collision for a particle with total energy $E$ as the moment where
  it hits an equipotential line of energy $E$.  We remark, however,
that for computational purposes (not carried out here) one should
define a collision differently, e.g., by a particle having a zero
velocity component perpendicular to an equipotential line. In order to
calculate $\tau_c$ for the soft Lorentz gas we straightforwardly adapt
the phase space argument of Eq.~(\ref{tauE}): Instead of considering
the length of the exits to calculate $\omega$ we replace them by the
length of the walls inside the trap. For this we use the arc length
that originates from the angle $\pi/3$ and the radius $r(E)$ of the
equipotential line determined by $V(r)=E$. There are three arcs
leading to $l=\pi r(E)$ and
$$\omega=\pi r(E)2v^2(E),$$ 
in analogy to Eq.~(\ref{omegaMZ}). 
The total volume of the phase space $\Omega$ is the same as in Eq.~(\ref{OmegaE}).
If an average constant velocity $v$ is assumed as before, then
\begin{equation}
\tau_c=\frac{A(W(E))}{r(E)v}.
\label{tauEB}
\end{equation}
Considering an average velocity given by $v_{\rm ave}=l_c/\tau_c$,
substituting $l_c=\tau_c v_{\rm ave}$ in Eq.~(\ref{D_b}) and using
Eq.~(\ref{tauEB}) leads to our final expression
\begin{equation}
D_{\rm B}(E)=\frac{A(W(E))}{4r(E)}v_{\rm ave}(E).
\label{boltzmanE}
\end{equation}

\subsection{Estimation of the velocities at the exit of a trap}

Generally the velocity of a particle in a soft Lorentz is not
constant, and it is not possible to obtain analytical forms for it.
Here we work out three approximations for the velocity when a particle leaves a trap.

\begin{enumerate}

\item We assume that a particle has a maximum velocity at the moment of exiting a trap.
Using energy conservation we find
$$v(E)=\sqrt{2(E-V(r))}.$$
The potential is given by Eq.~(\ref{fermi_pot}) at each point of the lattice.
Consider the base of a triangle in the array as the $x$-axis and let us take
into account the contributions of two adjacent potentials $V_1$ and
$V_2$, as shown to the right of Fig.~\ref{w_V1}.
For simplicity we continue our analysis by considering only the contribution of the 
potential on the $x$-axis, denoted as
\begin{equation}
V(x)=V_1(x)+V_2(x)=\frac{1}{1+\exp((|x|-r_o)/\sigma)}+\frac{1}{1+\exp(|x-L|-r_o)/\sigma)}.
\label{V1plusV21d}
\end{equation}
According to the right panel of Fig.~\ref{w_V1} the minimum of the
potential along the $x$-axis ($y=0$) is located in the middle of the
gap, $x_{\rm min}=r(E)+W(E)/2 $ or $x_{\rm min}=r_o+w/2$. The energy
threshold is then given by
$$E_*=V(r_o+w/2)=V_1(r_o+w/2)+V_2(r_o+w/2)=2/(1+\exp(w/(2\sigma))).$$
Correspondingly, the maximum velocity is
\begin{equation}
v_{\rm max}(E)=\sqrt{2(E-V(r_o+w/2))}.
\label{v_max}
\end{equation}
This is a function of the energy and independent of the gap size.

\item Alternatively we can define an average velocity according to an
  average potential in the gap.  Due to symmetry, we can
  calculate the integral of the potential $V_1(x)+V_2(x)$, again along
  the $x$-axis, in the first half of the region and average over
  $W(E)/2$,
$$V_{\rm ave }(W(E))=\frac{2}{W(E)}\int_{r(E)}^{r_o+w/2} ( V_1(x)+V_2(x) ) dx.$$
Plugging in the functional forms for $V_1$ and $V_2$ we obtain
\begin{equation}
V_{\rm ave}(E,W(E))=2+\frac{2\sigma}{W(E)}\ln\Bigg[ \frac{1+\exp(r(E)-r_0)/\sigma)}{1+\exp((L-r(E)-r_o)/\sigma)}\Bigg].
\label{AveIntV1plusV2}
\end{equation}
This yields an expression for the velocity that is a function of the energy and
the average potential $V_{\rm ave}$ at the exit of the trap,
\begin{equation}
v_{\rm ave}(E,W(E))=\sqrt{2(E-V_{\rm ave}(W(E))}.
\label{v_ave}
\end{equation}

\item  Finally, let us approximate the average velocity in a gap of size $W(E) $ by
\begin{equation}
v=\frac{2}{W(E)}\int_{r(E)}^{r(E)+W(E)/2} \sqrt{2(E-V(r))}dr,
\label{v_num}
\end{equation}
where $V(r)=V_1(r)+V_2(r)$.
This integral is not solvable analytically, but we can compute it numerically.
\end{enumerate}
Next, we compare our random walk approximations $D_{\rm MZ }(E)$ and
$D_{\rm B }(E)$ with our numerical results by using these three
estimates for the velocities .

\subsection{Comparison between random walk approximations and numerical results}

\begin{figure}
\centering
\begin{minipage}[c]{0.495\textwidth}
\centering 
\resizebox{\columnwidth}{!}{\rotatebox{-90}{\includegraphics{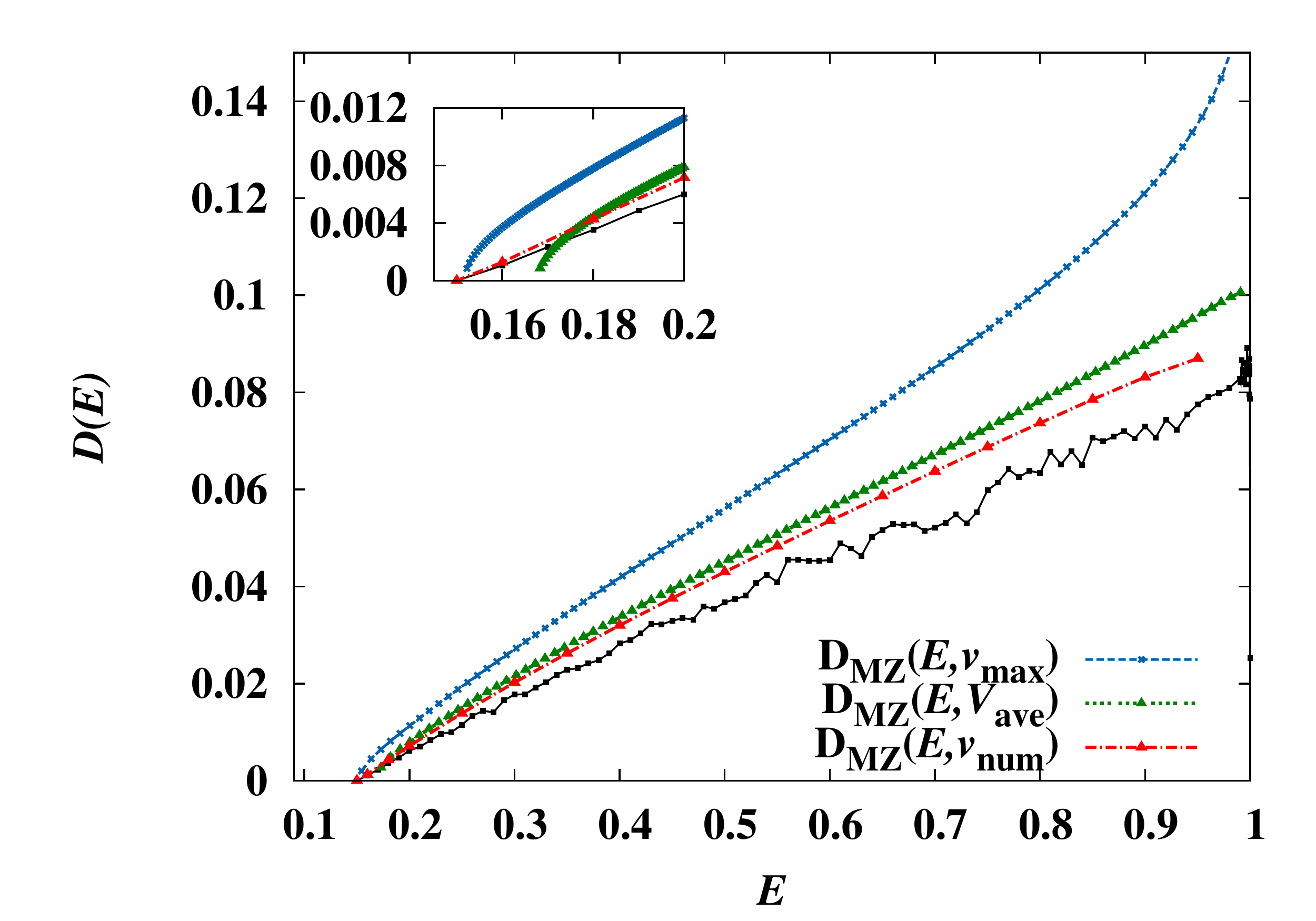}}}
\end{minipage}
\begin{minipage}[c]{0.495\textwidth}
\resizebox{\columnwidth}{!}{\rotatebox{-90}{\includegraphics{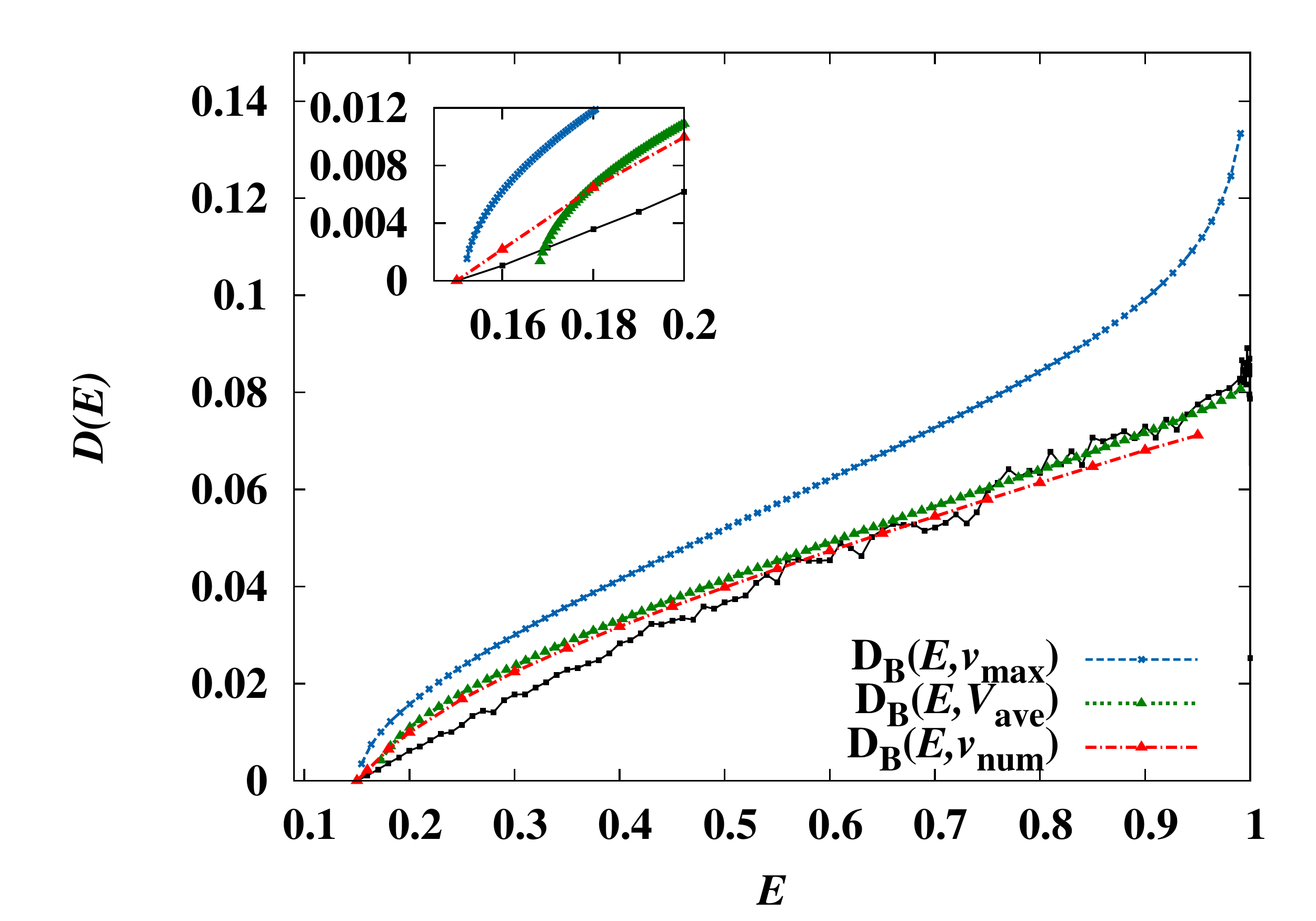}}}
\end{minipage}
\caption{Diffusion coefficient $D(E)$ as a function of the energy $E$
  in the small energy regime $E<1$.  The black (non-smooth) line shows
  the simulation results obtained from the same input parameters as in
  Fig~\ref{all}.  Left panel: Machta-Zwanzig random walk approximation
  $D_{\rm MZ}(E)$ given by Eq.~(\ref{D_rw}).  Right panel: Boltzmann
  random walk approximation $D_{\rm B}(E)$ given by
  Eq.~(\ref{boltzmanE}).  For each case three different approximations
  of the velocities have been used as described in the text.}
\label{ModelII_MZandB}
\end{figure}

The approximations based on $D_{\rm MZ}$ with different velocities are
compared to simulation results for the diffusion coefficient with
$E\leq 1$ in the left panel of Fig.~\ref{ModelII_MZandB}. As expected,
$D_{\rm MZ}$ with an average velocity does not reproduce the onset of
diffusion where $D=0$ while $D_{\rm MZ}$ that uses a maximum velocity
recovers the threshold (see the inset).  The approximation $D_{\rm
  MZ}$ with the maximum velocity is not accurate indicating that not
many particles travel with the maximum velocity but with some average
velocity instead.  According to the inset, $D_{\rm MZ}(E,v_{\rm num})$
captures the threshold and especially gives the correct asymptotics
when $E\to E_*$.

The approximations of $D_{\rm B}$ with three different velocities are
shown in the right panel of Fig.~\ref{ModelII_MZandB}.  The result
obtained by using the average velocity agrees well with the numerical
data at larger energies $E\to1$.  
Note that in this system we use a smoothness of $\sigma = 0.01$, which
produces a smooth steep potential similar in structure to the hard
Lorentz gas, at least for energies smaller than the maximum of the
potential.  Therefore particles are more likely to travel by ``free
flights'' before getting close to a scatterer.  This explains why
$D_{\rm B}$ is a better approximation than $D_{\rm MZ}$. However, our
Boltzmann approximation fails to reproduce $D(E)$ as the energy
approaches the onset of diffusion, since the approximation is based on
collisionless flights, and $\tau_c$ is defined even when $E<E_*$.
$D_{\rm B}$ still manages to capture the threshold with $v_{\rm max}$
and $v_{\rm ave}$, but the shapes of the curves do not match (see the
inset). We remark that the impact of varying the smoothness on
diffusion in the soft Lorentz gas has been investigated in
Refs.~\cite{KGSSR18,thesisGil}.


\section{Diffusion coefficient for intermediate energies}
\label{sec:transition}
Next we explore the regime of intermediate energies, i.e., $1<E<2$.
According to Fig.~\ref{all} the diffusion constant has a clear kink
with changing curvature at $E=1$ suggesting a mechanism of diffusion
that is different from the small energy regime as the energy passes
this value. A key feature of this regime is that at energies below
$E=1$ particles cannot pass the maxima of the potential but are
restricted to a certain area in configuration space, which we called
traps. However, when the total energy starts to exceed $E=1$ a
particle can cross all these maxima, which generates a novel regime of
diffusion.

\begin{figure}
\centering
\resizebox{0.65\columnwidth}{!}{\rotatebox{-90}{\includegraphics{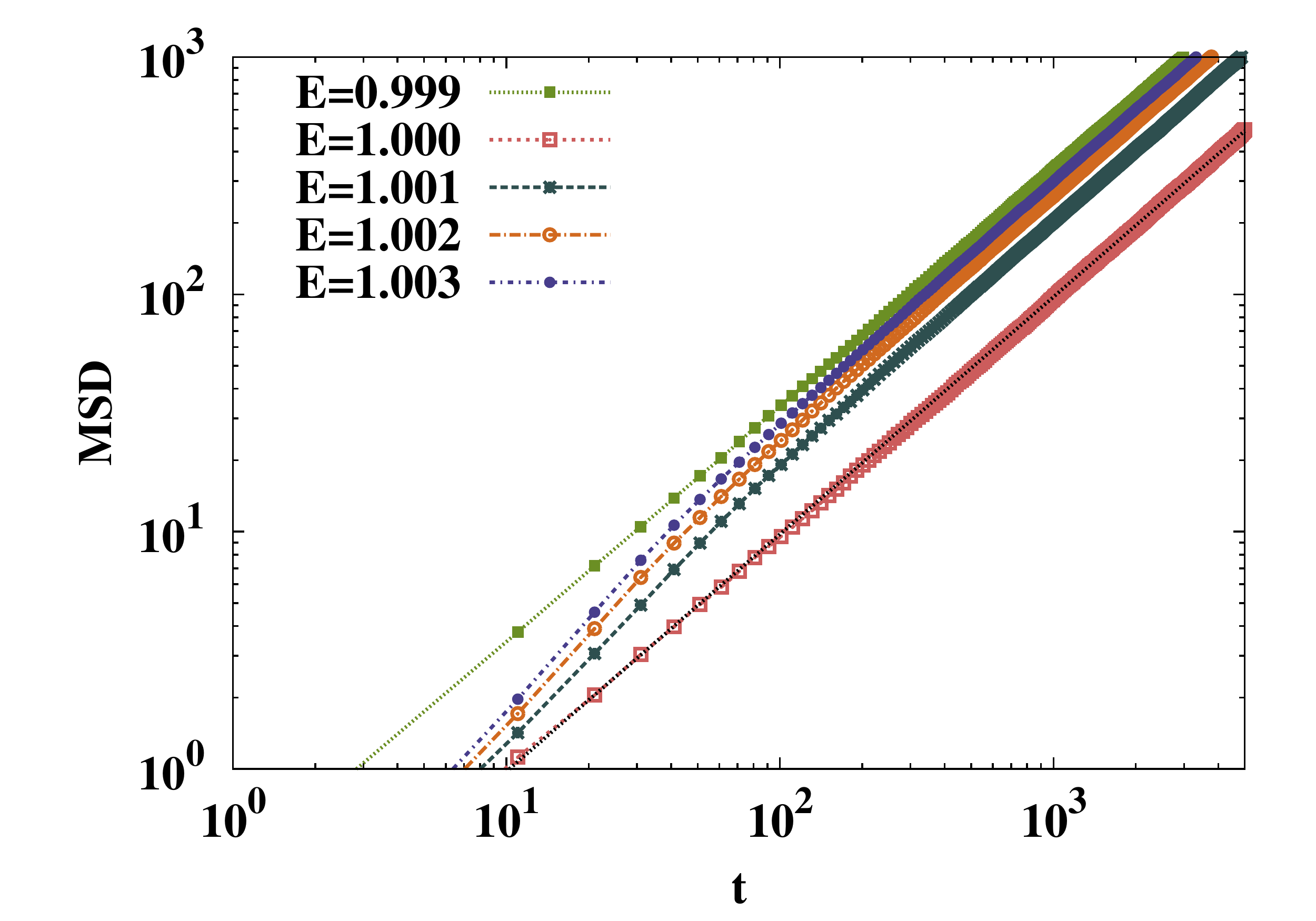}}}
\caption{Numerical results for the mean square displacement (MSD) as a
  function of time $t$ for different energies $E$ in the transition
  zone close to $E=1$. The input parameters are the same as in
  Fig.~\ref{all}.}
\label{MSDE1}
\end{figure}

In Fig.~\ref{MSDE1} we first show numerical results of the MSD as
a function of time $t$ around $E=1$.  In all cases the MSD shows normal
diffusion in the long time limit, although there are some features
that distinguish the regimes of $E<1$ from $E>1$. In particular, there
is a transient in the MSD at energies $E\geq 1.001$, which is not
present at $E\leq 1$.
Note that at $E=1+\epsilon$, $\epsilon>0$, the area available in the
configuration space becomes unbounded, that is, particles with
energies $E>1$ are allowed to travel everywhere, even in regions very
close to the centre of a scatterer. When reaching this position,
however, particles have very small but non-zero kinetic energy. This
causes an unstable trapping mechanism, which suppresses diffusion when
$E\to1$.

We can now construct a random walk approximation for diffusion in this
regime based on traps of slow motion.  For this purpose, we need to
redefine a trap {\em on a scatterer} and also to compute the average
rate at which a particle leaves this trap. It is not straightforward
to exactly determine the area where this trapping occurs, since this
depends on how one defines ``slow motion''. The potential is radially
symmetric at each lattice point, hence let us assume that the trapping
mechanism takes place on a circle with some radius $r_s$ centered at
each scatterer.  According to the definition of our Fermi-type
scatterer, $\sigma \to 0$ corresponds to enlarging the top of the
potential. Let us take a constant value $r_s=1=r_o$ as our first
approximation for the radius of the circle.

\begin{figure}
\centering
\begin{minipage}[c]{0.495\textwidth}
\resizebox{\columnwidth}{!}{\includegraphics{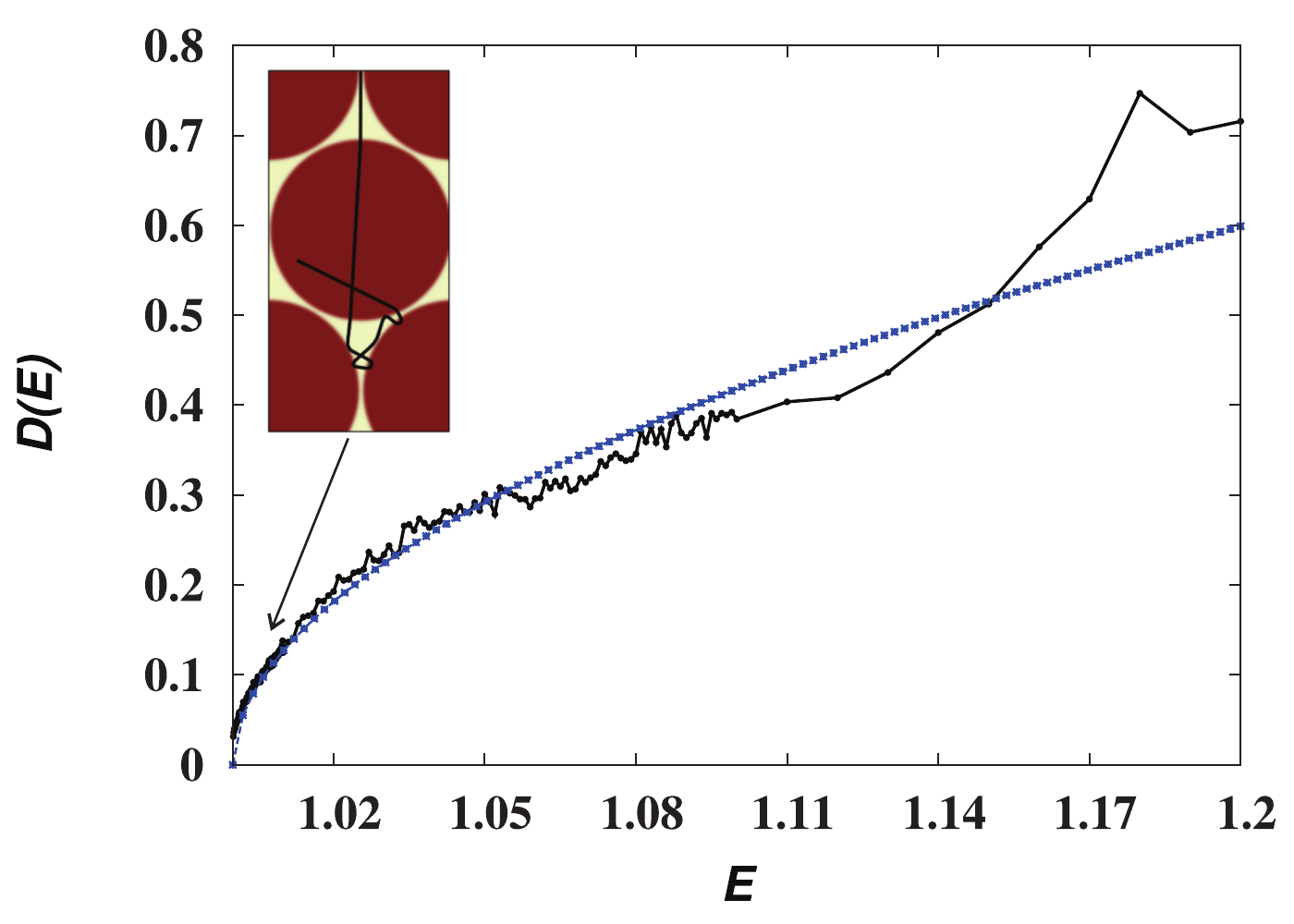}}
\end{minipage}
\begin{minipage}[c]{0.495\textwidth}
\resizebox{\columnwidth}{!}{\includegraphics{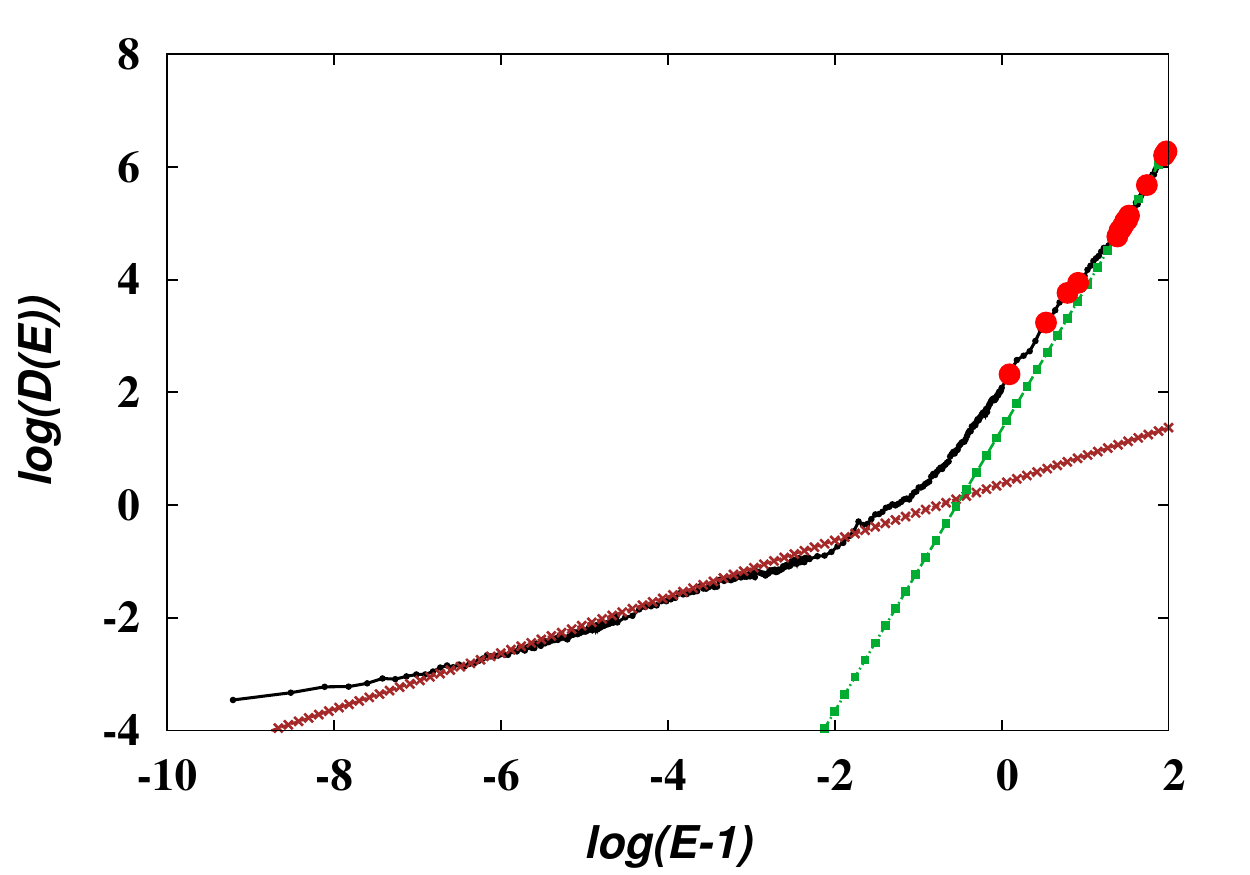}}
\end{minipage}
\caption{ Diffusion coefficient $D(E)$ in the transition regime $ E\to
  1^+$ obtained from simulations (black dotted lines) together with
  our random walk approximation in this regime
  Eq.~(\ref{D_E_transition}). Results are given both on linear (left
  panel) and log-log scales (right panel). In the right panel the red
  line has a slope $1/2$, the green line has slope $5/2$, and red
  circles indicate the presence of islands of stability in phase
  space. The inset in the left panel shows a typical trajectory in
  this diffusive regime.}
\label{transitionModelII}
\end{figure}

We first need to calculate the velocity of a particle in this region.
In order to do so, we assume that inside the circle, or trap, the
velocity is constant. The potential energy is (close to) maximal in
this circle, therefore the kinetic energy as well as the velocity are
(close to) minimal. Hence, $v_{\rm min}=\sqrt{2(E-V_{\rm max})}$.
The total phase space of the trap is $\Omega=A_{\bullet}2\pi v$, where
$A_{\bullet}(r_{\rm s})$ is the area of the circle with radius $r_s$,
and $v$ is some constant minimal velocity.  Then the available phase
space where particles leave the trap in time $\Delta t$ is
$$\omega =C  2 v^2,$$
where $v$ is the constant velocity defined above and $C$ is the length
of the available portion of the trap where particles escape, i.e., the
length of the circumference with radius $r_s$.  Using Eq.~(\ref{tauE})
we obtain the escape time as
\begin{equation}
\tau(E)=\frac{A_{\bullet}(r_{\rm s}) }{2r_{\rm s} v_{\rm min} }.
\label{tauSlowTrap}
\end{equation}
We can now substitute Eq.~(\ref{tauSlowTrap}) into the random walk
approximation for $D$ Eq.~(\ref{D_MZ}), where $l$ is considered to be
the distance between two traps or centres of scatteres, which in this
case is given simply by the geometry of the problem as $l=L=2r_o+w$.
Finally we get
\begin{equation}
D_{\rm s}(E)=\frac{r_o(2r_o+w)^2}{2  A_{\bullet}(r_{\rm s})}v_{\rm min}(E).
\label{D_E_transition}
\end{equation}
In this approximation we see that
$$D_{\rm s}(E)={\rm const.}\cdot v_{\rm min}(E) = {\rm const.} \sqrt{2(E-V_{\rm max})}.$$
Choosing a constant potential $V=1$ this expression yields
\begin{equation}
D_{\rm s}(E)\sim \sqrt{E-1}.
\label{sqrt_Eminuns_one}
\end{equation}
Let us now compare our numerical results to this approximation when $E\to 1^+.$ 
In the left panel of Fig.~\ref{transitionModelII} we show the numerically obtained
diffusion coefficient in the transition regime together with a numerical fit to the data
of the form
\begin{equation}
 D(E)=a(E-1)^b.
\label{D_slowTrap}
\end{equation}
With this fit we find an exponent of $b=0.518$ in the energy interval
(1.0001, 1.2). This agrees quite well with the analytical description
of Eq.~(\ref{D_E_transition}). Our approximation makes assumptions on
the radius $r_s$ and does not match exactly to the shape of $D(E)$.
On the other hand, it reveals the asymptotic form $\sqrt{E-1}$ as
$E\to 1^+$.  This implies that the motion of particles is dominated by
slow motion close to the centres of the scatterers leading to a full
suppression of $D(E)$. To see this more clearly, the right panel of
Fig.~\ref{transitionModelII} shows $\log D(E)$ as a function of $\log
(E-1)$. We observe a regime where the data matches the red
line with slope $m=1/2$, equivalent to the square-root behavior in the
linear graph in the left panel. In addition, we find the transition to
a different regime where $m=5/2$ (green line), which we discuss in
Sec.~\ref{HERnumerics}.

At small values $E\to 1^+$ there is a discrepancy between the
simulation data and the randow walk approximation.  This is a delicate
regime in terms of numerical accuracy. Moreover, our approximation
assumes that $v$ is constant in a circle with radius $r_s$ of area
$A_{\bullet}$. While in the approximation there is an abrupt jump at
$r_{\rm s}$, the potential that we use for the simulations is a smooth
function of the position.  Performing numerical simulations with
higher precision in this regime and modelling $V_{\rm max}$ as
accurately as possible could reduce the difference between the
approximation and the data enhanced by the logarithmic scale.

The functional form of $D(E)$ reflects a dynamics where particles
travel with small velocity over the top of the scatterers and
randomize in the middle of the zones with minimum potential energy,
see the inset of Fig.~\ref{transitionModelII}. This is supplemented by
a contribution from particles that travel in straight lines until they
get scattered. The last effect is more pronounced as the energy
increases, as we discuss in the following. 

\section{Diffusion coefficient for high energies}
\label{HERnumerics}
\begin{figure}
\centering
\begin{minipage}[c]{0.48\textwidth}
\resizebox{\columnwidth}{!}{\includegraphics{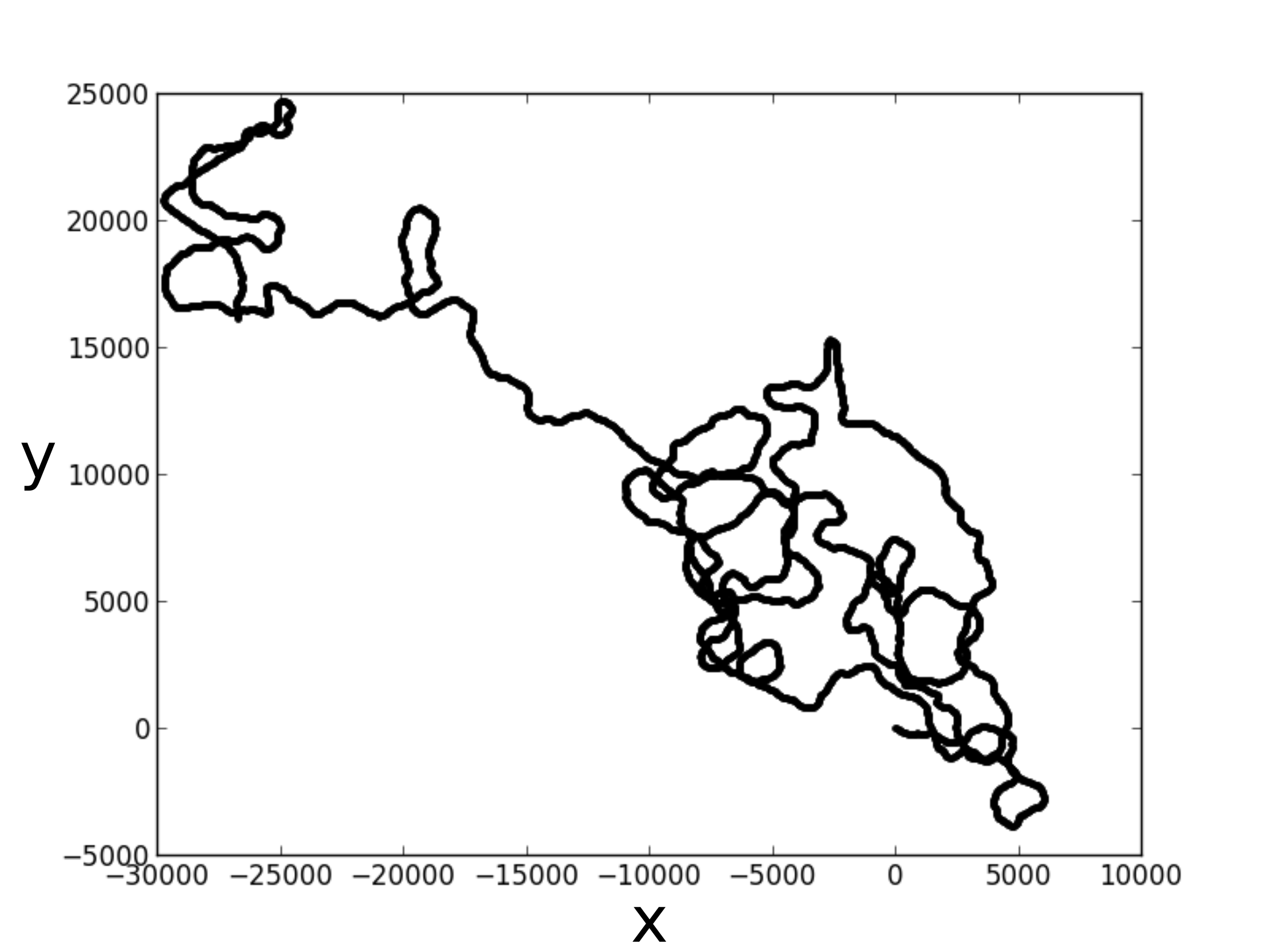}}
\end{minipage}
\begin{minipage}[c]{0.48\textwidth}
\resizebox{\columnwidth}{!}{\rotatebox{-90}{\includegraphics
{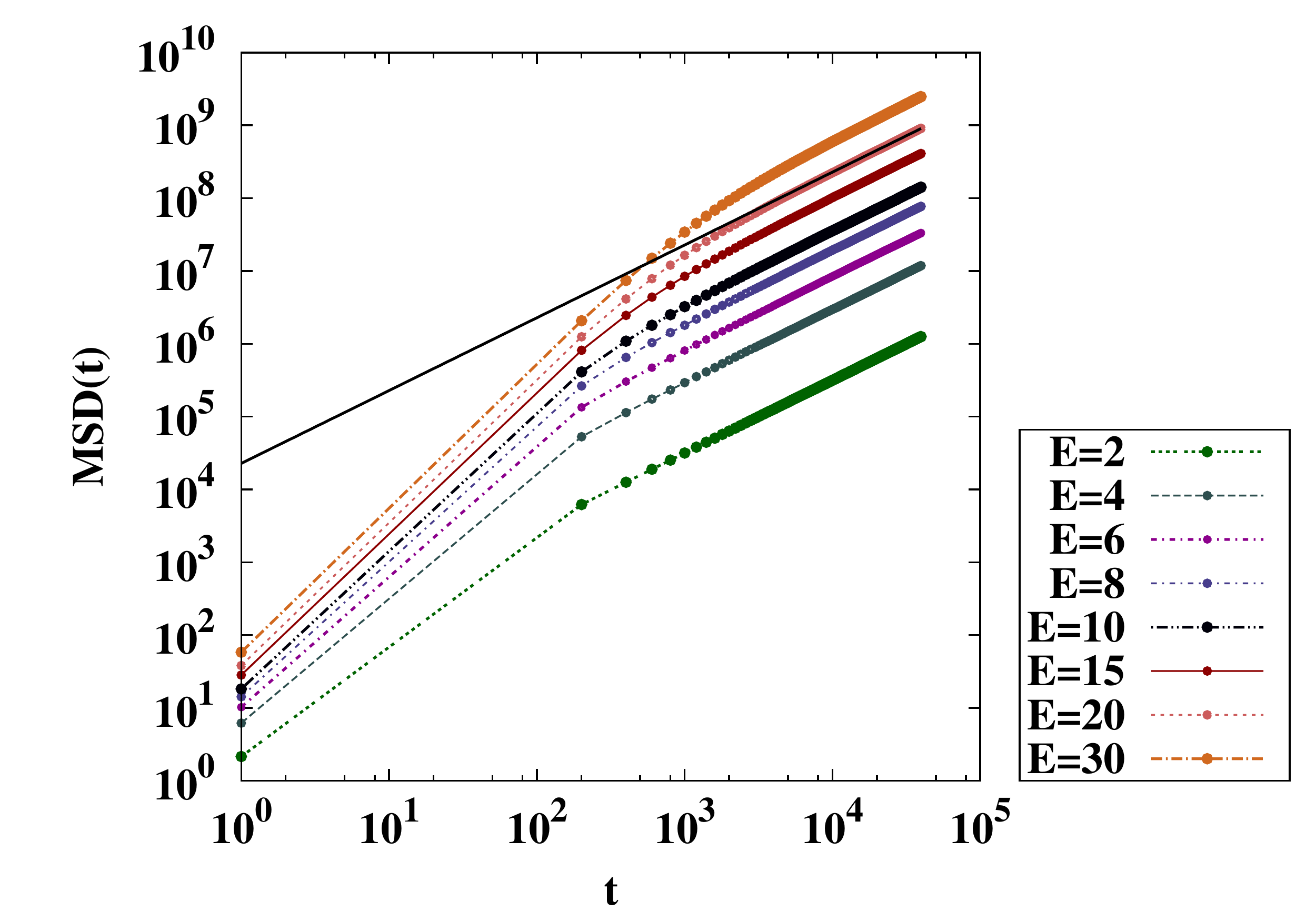}}}
\end{minipage}

\caption{Left panel: Example of a trajectory at $E=40$ with iteration
  time $t=40000$.} Right panel: Time dependence of the mean square
  displacement (MSD) in the high energy regime. The black line
  indicates a slope of one.  
\label{MSD_largeE}
\end{figure}

Our final quest is to characterize the diffusion coefficient $D(E)$
for large energies.  Here we compare our numerical results to
theoretical approximations put forward by Aguer et
al.~\cite{aguer2010elastic} and Nobbe~\cite{Nob95}.  For this purpose
we need to make sure that we deal with an interval of the energy
parameter that exhibits normal diffusion.  To illustrate the
  problem of identifying such energies, we show a typical trajectory
  at high energies in the left panel of Fig.~\ref{MSD_largeE}: We see
  that the particle travels in the same direction for long periods of
  time by then changing its direction, it repeats this and in the long
  run generates a random pattern at large scale. This corresponds to
  the fact that as the energy parameter is increased the particles
  move faster and consequently travel long distances before changing
  their direction, meaning the randomization process takes longer than
  in the small energy regime. But this purely qualitative picture does
  not guarantee that normal diffusion exists for this or other energy
  parameters.  Quantitatively, the MSD should tell whether there is
  normal diffusion, which is what we explore next.

In the right panel of Fig.~\ref{MSD_largeE} we show the MSD for
  energies $2\le E\le30$, which roughly corresponds to the high energy
  regime of diffusion shown in Fig.~\ref{all}.  The black line
  indicates a slope equal to one yielding normal diffusion, where
  according to Eq.~(\ref{diffusion}) the diffusion coefficient $D(E)$
  exists. For all these energies the MSD eventually appears to grow
  linearly in time in the long-time limit yielding $D(E)$ plotted in
  Fig.~\ref{all}. However, we see that for higher energies the
  transition time to the onset of the linear asymptotic regime shifts
  to considerably longer times: While for $E=2$ the MSD appears to be
  linear starting from around $t\sim10^2$, for $E=30$ values of $t$
  larger than $10^3$ are required. It is thus not entirely clear
  whether there is a linear asymptotic regime for arbitrarily high
  energies, or whether there exists an energy value starting from
  which a diffusion coefficient does not exist anymore. Secondly, we
  can not tell whether the trend of linear growth in time really
  continues at longer simulation times. That the situation is indeed
  more subtle becomes clear by looking in detail at the phase space of
  our model, which gives us a more precise method than simply plotting
  the MSD to decide about normal diffusion.

    Our aim is to check for periodic islands of stability in phase
  space, which are known to have a major impact on transport
  properties \cite{ZGNR86,Zas02,KRS08}. This holds in particular for
  so-called {\em accelerator mode islands}
  \cite{LiLi92,MaRo14,HaCo18}, which in our model originate from
  quasi-ballistic (q-b) trajectories.  With a q-b trajectory we mean
  the path of a particle that travels on average in one direction
  across the lattice, that is, the path wiggles periodically around a
  straight line, where the wiggles are due to specific microscopic
  scattering events; see Fig.~3 in Ref.~\cite{KGSSR18} for explicit
  examples and Ref.~\cite{thesisGil} for full details. But even
  initial conditions outside these stable component that `stick' to
  the boundary of an island can have an effect on averages of
  observables \cite{RKZ99,HaCo18}.  Therefore, if we find islands of
  stability due to q-b trajectories in configuration space we cannot
  expect to have normal diffusion in the long time limit
  \cite{ZGNR86,Zas02,KRS08,LiLi92}.  For this reason we need to
  exclude any parameter values where we find q-b periodic orbits from
  an analysis of an energy-dependent diffusion coefficient $D(E)$.

    In order to find islands of stability in phase space we use the
approach via Poincar\'{e} surfaces of section (PSoS) as described in
Refs.~\cite{KGSSR18,thesisGil}.  In our case the PSoS is defined by
the plane $(x,\sin \theta)$, where $x$ is the position of a particle
when it leaves the rhomboid unit cell (see Fig.~\ref{w_V1}), and as
before $\theta$ is the angle between its velocity vector and the
normal to the boundary. As a constraining numerical factor a given
ensemble of initial conditions does not necessarily catch tiny islands
of stability: The smaller the island, the more difficult it is to
detect it numerically, as it only appears if we choose initial
conditions that are in the island. The islands of stability that we
find are indeed very tiny, and they correspond to two different types
of periodic motion: There are islands related to q-b trajectories, as
discussed above, but also other islands that display localized motion.
While the latter do not yield anomalous diffusion, the q-b
trajectories do \cite{ZGNR86,Zas02,KRS08,LiLi92,MaRo14,HaCo18}.
Islands of stability exist for many energy values in our model, in
particular, trivial islands generated by the main symmetry channels
due to the focusing of the potential walls, and other islands where
particles fly over the top of the scatterers.

\begin{figure}
\centering
\begin{minipage}[c]{0.489\textwidth}
\centering 
\resizebox{\columnwidth}{!}{\rotatebox{-90}{
\includegraphics{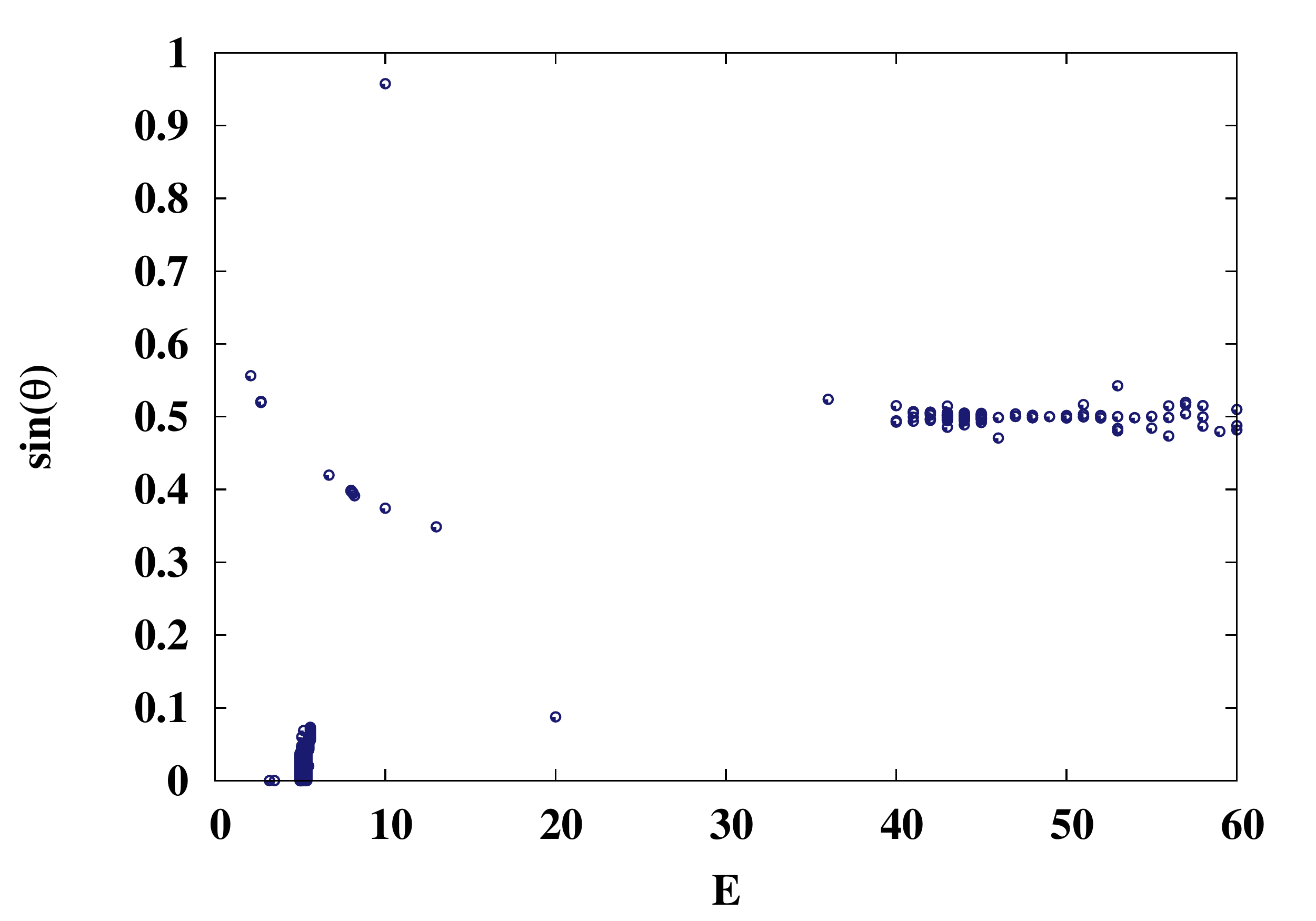}}}
\end{minipage}
\begin{minipage}[c]{0.489\textwidth}
\centering 
\resizebox{\columnwidth}{!}{{
\includegraphics{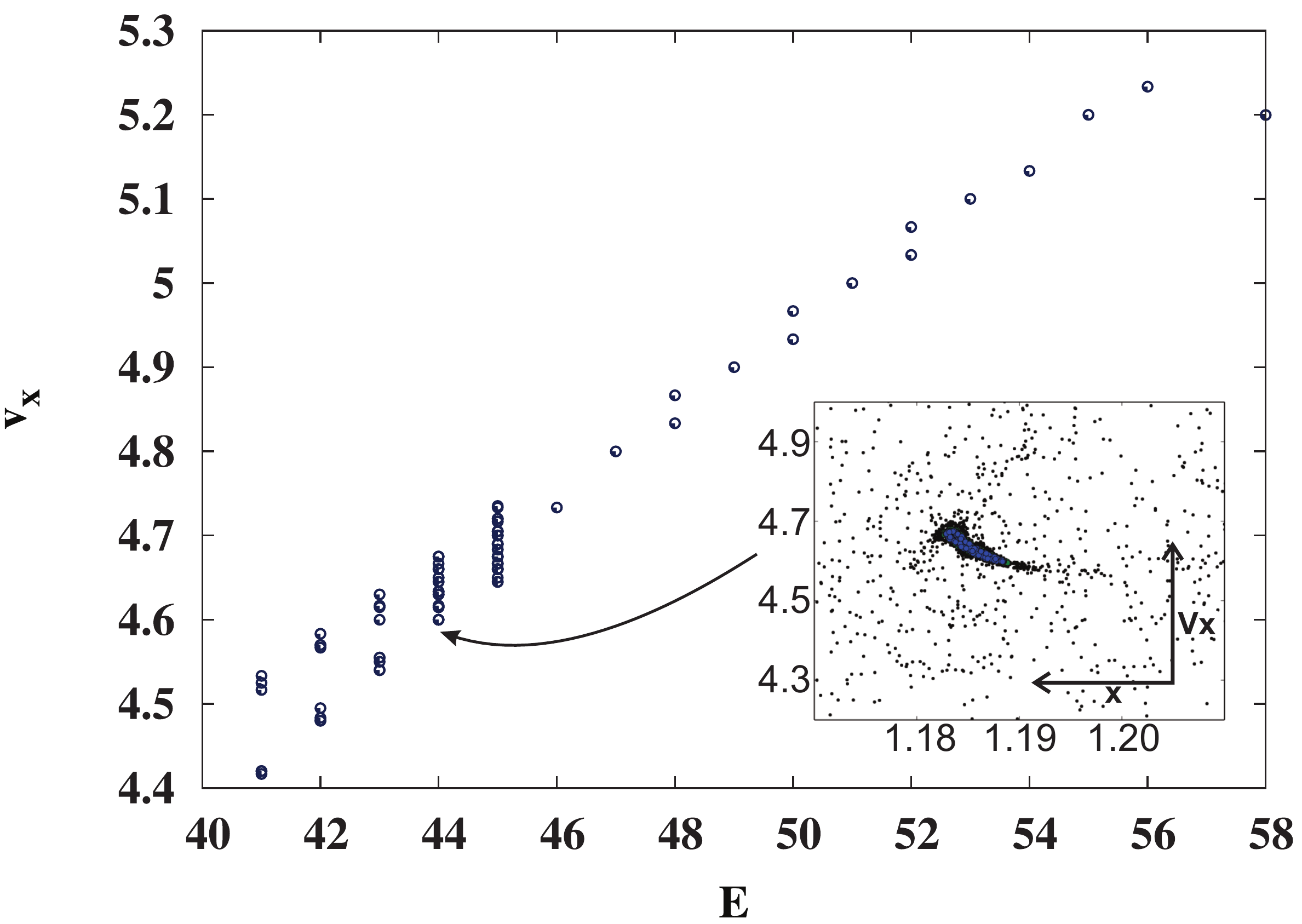}}}
\end{minipage}
\caption{ Islands of stability in phase space: Each circle in both plots represents a
  potentially stable island in phase space for a given energy parameter $E$.
  Left panel: The $y$-axis indicates the sine of the angle $\theta$
  between the velocity vector at the moment of exiting the gap and the
  normal to the boundary.  Right panel: As in the left panel but with
  the velocity component $v_x$, displaying a backward bifurcation.
  Inset: Stability island due to a quasi-ballistic trajectory at
  $E=44$.}
\label{diagramE}
\end{figure}

      Our results are summarised in Fig.~\ref{diagramE}, which shows
  parameter values where we detect periodic islands of stability. The
  shapes of their corresponding trajectories vary depending on the
  energy. The right panel of Fig.~\ref{diagramE} depicts a region of
  the PSoS at $E=44$ where a stability island is found. The coordinate
  space trajectories corresponding to this island are almost straight
  lines, as can be expected due to the high total energy, hence they
  are not shown. As accordingly the diffusion coefficient $D(E)$ does
  not exist at these energies, we have plotted red filled circles
  between respective values of the diffusion coefficient curve in
  Fig.~\ref{all}.  A more detailed analysis \cite{thesisGil} reveals
  that the structure of the phase space is rich, even for smaller
  energies.

\begin{figure}
\centering
\resizebox{0.6\columnwidth}{!}{{
\includegraphics{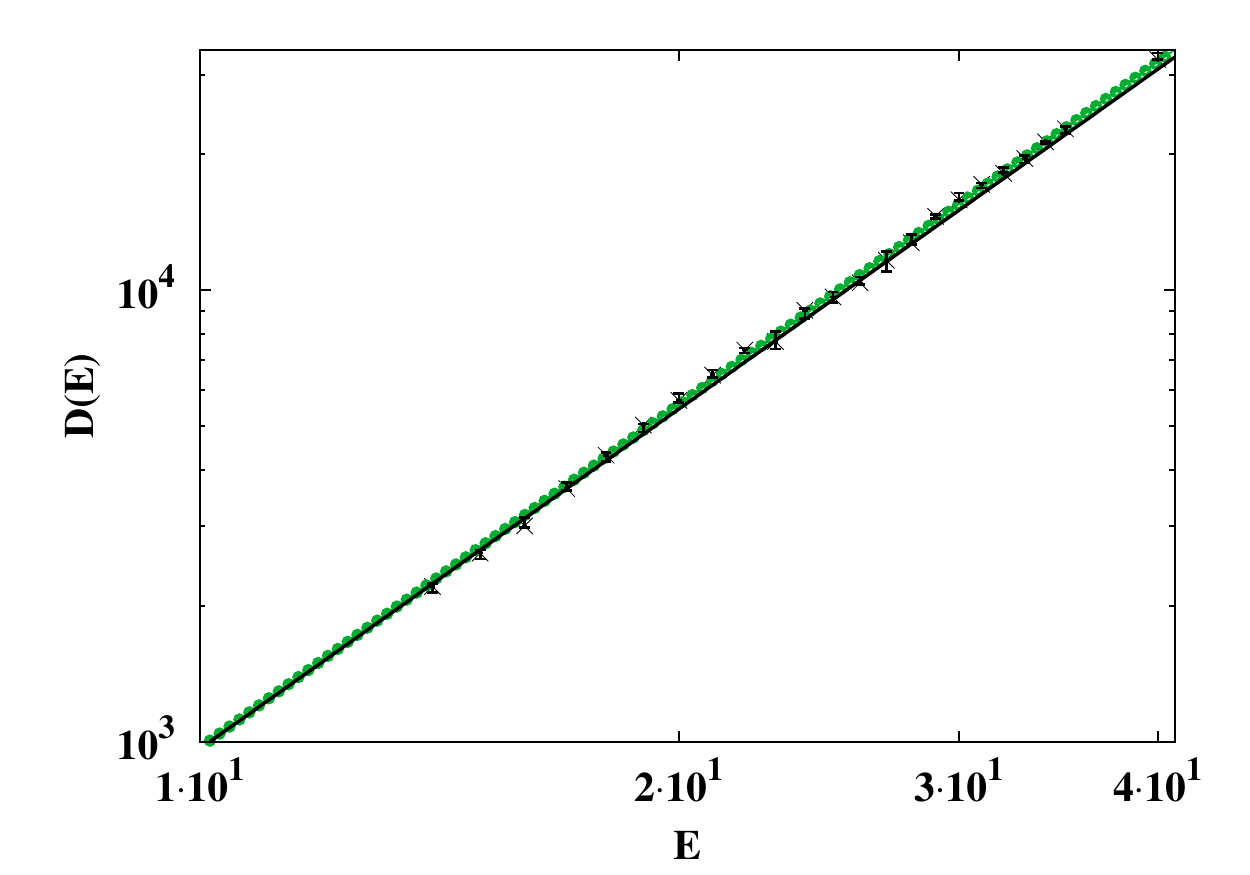}}}
\caption{Diffusion coefficient $D(E)$ obtained from simulations as a
  function of the energy $E$. The numerical data is the same as in
  Fig.~\ref{all}.  The green line indicates a slope of $m=5/2$ as
  predicted in Ref.~\cite{aguer2010elastic}.}
    \label{fig:image}
\end{figure}

To find the normal diffusion coefficient $D(E)$ as a function of the
energy, Fig.~\ref{diagramE} suggests to consider an interval of
energies $14<E<19$ or $21<E<40$. If we ignore the peak at $E=20$ we
can choose $14<E<40$.  Figure~\ref{fig:image} shows numerical results
on a double-logarithmic scale together with an error estimate; for
details of this estimate see Ref.~\cite{thesisGil}. The error bars
are relatively small. The green line indicates a slope of $m=5/2$ as
predicted in Ref.~\cite{aguer2010elastic}.
Fitting a function of the form $aE^m$ to the numerical data in the
intervals of energy $(21,35)$ yields $m=2.54$.  More fits and their
resulting exponents in different energy regimes, together with error
estimates, are presented in Table~\ref{table:tableDE}.

In Ref.~\cite{aguer2010elastic} it was concluded that at sufficiently
large energies $D(E)\propto E^{5/2}$ while Ref.~\cite{Nob95} which,
however, applies to attractive potentials, yields $D(E)\propto
E^{3/2}$.  Overall, the behaviour detected here seems to be very close
to the one predicted by the former theory but very different from the
latter. By observations of trajectories in configuration space (see
the right panel of Fig.~\ref{MSD_largeE}), our model behaves similarly
to the dynamics underlying the model in Ref.~\cite{aguer2010elastic},
where particles travel for long times before changing the direction by
a small angle.  In contrast, for the model in Ref.~\cite{Nob95}
particles travel over short distances and change the direction of the
path by a large angle. This difference was also noted in
Ref.~\cite{aguer2010elastic}. We note that the simulations in the
latter work always yielded normal diffusion even for large energies.
This is due to the model used therein, where there is no mechanism of
generating islands in phase space.
\begin{table}
\centering
\caption{Exponent $m$ in $D(E)=aE^m$ obtained from fits to the 
simulation data over different intervals of energies $E$. The last column indicates the error of the fits.}
\label{table:tableDE}       
\begin{tabular}{lrr}
\hline\noalign{\smallskip}
Energy &    $m$ & $+/-$ \\ [1ex] 
\noalign{\smallskip}\hline\noalign{\smallskip}
$2<E<35$         & 2.524 &$0.0188 $\\
$14<E<40$        & 2.540 &$ 0.0186$\\
$21<E<35$        & 2.521 &$0.0376$\\ 
$21<E<40$        & 2.545 &0.0257\\ 
\noalign{\smallskip}\hline
\end{tabular}
\end{table}
\section{Conclusions} 
\label{sec:conclusions}
We have studied diffusion in a soft periodic Lorentz gas in which the
hard walls of the conventional Lorentz gas scatterers were replaced by
repulsive Fermi potentials. Our goal was to understand how diffusion
depends on varying the energy as a control parameter in this system.
For this purpose we computed the diffusion coefficient as a function
of the total energy of a particle from simulations. We compared our
numerical results with simple analytical random walk approximations.

We distinguished three different diffusive regimes: (1) For small
energies, i.e., when the energy is less than the maximum of the Fermi
potential, there is an onset of diffusion which is well approximated
by a random walk approximation put forward by Machta and Zwanzig
\cite{MaZw83} based on a simple phase space argument. For slightly
larger energies the energy dependence of the diffusion coefficient is
well explained by a Boltzmann-type random walk approximation which
employs a collisionless flight argument. (2) There is a specific value
of the energy at which a particle can for the first time travel over
the top of a Fermi potential. This defines the onset of a second
diffusive regime, where a particle is getting trapped on the top of
each potential. In this regime we observe a full suppression of
diffusion with a square root dependence on the energy, as is explained
by another simple random walk argument similar to the one of Machta
and Zwanzig. (3) For large energies the energy-dependent diffusion
coefficient yields a power law $D(E)\propto E^{5/2}$ in agreement with
the random walk argument presented in Ref.~\cite{aguer2010elastic}. On
top of this dynamics there are parameter regions exhibiting
superdiffusion which, however, were not the focus of our present
study.
	
We remark that we have also studied energy-dependent diffusion in this
soft Lorentz gas with a second, different setting of parameters
modeling a shallower potential. Here we observed that for small
energies the Boltzmann approximation is in general a better
approximation outperforming the one by Machta and Zwanzig for all
energies even at the onset of diffusion \cite{thesisGil}. However, for
these parameters q-b islands start to appear even at smaller energies.
Our numerical results furthermore suggest that islands of stability in
phase space are ubiquitous for large enough energies, irrespective of
particular values of the other parameters. We also observed that at
large enough energies the size of the islands increases with the
energy, in agreement with Ref.~\cite{Pan}.  This will profoundly
obscure the underlying dependence of the diffusion coefficient as a
function of the energy in this regime.

An interesting open question is to determine the precise type of
superdiffusion for parameter regions with islands in phase space by
matching the simulation data to predictions of a stochastic model,
such as L\'evy walks \cite{CGLS14b}. Secondly, it would be exciting if
these different diffusive regimes could be detected in experiments,
e.g., of diffusion in molecular graphene. Here perhaps the temperature
of the system could be varied mimicking our variation of the total
energy of a particle. Especially the suppression of diffusion at
intermediate energies should be a phenomenon that would be interesting
to be observed in experiments.\\

\small{
S.G.G.\ acknowledges support from the Mexican National Council for Science
and Technology (CONACyT) by scholarship no.~262481.
R.K.\ thanks Prof.~Krug from the U.\ of Cologne
and Profs.~Klapp and Stark from the TU Berlin for hospitality as a
guest scientist as well as the Office of Naval Research Global for
financial support. He also acknowledges funding from the
London Mathematical Laboratory, where he is an External Fellow.}\\

\small{
The first and the second author designed the research. All authors performed the research, but the first author contributed the majority to it. The computer code (published separately) was provided by the team around the
last author. All authors analysed the data, The first author wrote the manuscript, supported by all others.}


\begin{thebibliography}{}
\bibitem{Do99}
J.R. Dorfman, \textit{An introduction to chaos in nonequilibrium statistical mechanics} (Cambridge University Press, Cambridge, 1999)
\bibitem{Gas98}
P. Gaspard, \textit{Chaos, scattering, and statistical mechanics} (Cambridge University Press, Cambridge, 1998)
\bibitem{Kla07}
R. Klages, \textit{Microscopic chaos, fractals and transport in nonequilibrium statistical mechanics} (World Scientific, Singapore, 2007)
\bibitem{KlaDor}
R. Klages, J.R.Dorfman, Phys. Rev. Lett. \textbf{74}, 387-390 (1995)
\bibitem{KlDo99}
R.~Klages and {J.R.} Dorfman, Phys. Rev. E {\bf 59}, 5361-5383 (1999)
\bibitem{GrKl02}
J.~Groeneveld and R.~Klages, J. Stat. Phys. {\bf 109}, 821-861 (2002)
\bibitem{Chir79}
B. V. Chirikov, Phys. Rep. \textbf{52}, 263-379 (1979) 
\bibitem{RRW81}
A.B. Rechester, M.N. Rosenbluth and R.B. White, Phys. Rev. A {\bf 23}, 2664-2672 (1981)
\bibitem{BlGJ96}
J.A. Blackburn, N. Gr{\o}nbech-Jensen, Phys. Rev. E \textit{53}, 3068-3072 (1996) 
\bibitem{CGMR15}
F. Cagnetta, G. Gonnella, A. Mossa, S. Ruffo,  Europhys. Lett. \textbf{111}, 10002 (2015) 
\bibitem{ZGNR86}
A.~Zacherl, T.~Geisel, J.~Nierwetberg, and G.~Radons,
Phys. Lett. {\bf 114A}, 317-321 (1986)
\bibitem{MaRo14}
T.~Manos and M.~Robnik, Phys. Rev. E {\bf 89}, 022905 (2014)
\bibitem{Zas02}
G. Zaslavsky, Phys. Rep. \textbf{371} 461 (2002).
\bibitem{KRS08}
R.~Klages, G.~Radons, and I.M. Sokolov (editors), {\em Anomalous transport: Foundations and Applications}
(Wiley-VCH, Berlin, 2008)
\bibitem{Sza00}
D.~Szasz (editor), {\em Hard-ball systems and the Lorentz gas} (Springer, Berlin, 2000)
\bibitem{Lo05}
H.A. Lorentz, Acad. Amst., \textbf{7}, 438-453 (1905) 
\bibitem{BS81}
L.A. Bunimovich, Ya. Sinai,  Commun. Math. Phys. \textbf{78}, 247-280  (1980)  
\bibitem{GN90}
P. Gaspard, G. Nicolis,  Phys. Rev. Lett. \textbf{65}, 1693-1696 (1990) 
\bibitem{KlDe00}
R.~Klages and C.~Dellago, J. Stat. Phys. {\bf 101}, 145-159 (2000)
\bibitem{HaKlGa02}
T. Harayama, R. Klages, P. Gaspard, Phys. Rev. E \textbf{66}, 026211 (2002) 
\bibitem{KlKo02}
R.~Klages and N.~Korabel, J. Phys. A: Math. Gen. {\bf 35}, 4823--4836 (2002)
\bibitem{Det14}
C. Dettmann, Commun. Theor. Phys. \textbf{62}, 521 (2014) 
\bibitem{HaGa01}
T. Harayama and P. Gaspard, Phys. Rev. E, \textbf{64}, 036215 (2001) 
\bibitem{MaKl03}
L.~M{\'a}ty{\'a}s and R.~Klages, Physica D {\bf 187}, 165 (2004)
\bibitem{GaKl}
P. Gaspard and R. Klages, Chaos, \textbf{8}, 409-423 (1998) 
\bibitem{TurRom98}
D. Turaev and V. Rom-Kedar, Nonlin. \textbf{11}, 575 (1998)
\bibitem{RomTur99}
V. Rom-Kedar, D. Turaev, Physica D \textbf{130}, 187 (1999) 
\bibitem{LiLi92}
A.J. Lichtenberg and M.A. Lieberman, {\em Regular and chaotic dynamics}, 2nd Edition (Springer, New York, 1992)
\bibitem{KFAD04}
A. Kaplan, N. Friedman, M. Andersen, N. Davidson, Physica D \textbf{187}, 136145 (2004) 
\bibitem{Kna87}
A. Knauf, Commun. Math. Phys. \textbf{110}, 89-112 (1987) 
\bibitem{Nob95}
B. Nobbe, J. Stat. Phys. \textbf{78}, 1591-1605 (1995) 
\bibitem{aguer2010elastic}
B. Aguer, S. De Bievre, J. Phys. A: Math. Theor. \textbf{43(47)}, 474001 (2010) 
\bibitem{BZM85}
B. Bagchi, R. Zwanzig, M. C. Marchetti, Phys. Rev. A, \textbf{31}, 892-896 (1985) 
\bibitem{GZR87}
T. Geisel, A. Zacherl, G. Radons. Phys. Rev. Lett., \textbf{59}, 2503-2507 (1987) 
\bibitem{GZR88}
T. Geisel, A. Zacherl, G. Radons, Z. Phys. B, \textbf{71}, 117-127 (1988) 
\bibitem{GWNO90}
T. Geisel, J. Wagenhuber, P. Niebauer, G. Obermair, Phys. Rev. Lett. \textbf{64}, 1581-1584 (1990) 
\bibitem{Pan}
N. C. Panoiu, Chaos \textbf{10}, 166-179 (2000) 
\bibitem{YaZh}
J. Yang, H. Zhao, J. Stat. Mech.: Theor. Exp. \textbf{12}, L12001 (2010) 
\bibitem{Bal88}
P.R. Baldwin, Physica D \textbf{29}, 321 (1988)
\bibitem{LKP91}
A.~Lorke, J.P. Kotthaus, and K.~Ploog, Phys. Rev. B {\bf 44}, 3447-3450 (1991)
\bibitem{Weis91}
D.~Weiss, M.L. Roukes, A.~Menschig, P.~Grambow, K.~von Klitzing, 
  G.~Weimann, Phys. Rev. Lett. {\bf 66}, 2790--2793 (1991)
\bibitem{FGK92}
R. Fleischmann, T. Geisel, R. Ketzmerick, Phys. Rev. Lett. {\bf 68}, 1367--1370 (1992)
\bibitem{FlSS96}
M. Flie{\ss}er, G.J.O. Schmidt. H. Spohn, Phys. Rev. E {\bf 53}, 5690--5697 (1996)
\bibitem{Gibertini}
M. Gibertini, A. Singha, V. Pellegrini, M. Polini, G. Vignale, A. Pinczuk, L. N. Pfeiffer, K. W. West, Phys.
Rev. B \textbf{79}, 241406 (2009).
\bibitem{Rasanen}
E. R{\"a}s{\"a}nen, C. A. Rozzi, S. Pittalis, G. Vignale, Phys. Rev. Lett. \textbf{108}, 246803 (2012).
\bibitem{GMKGM12}
K.~K. Gomes, W. Mar, W. Ko, F. Guinea, H. C. Manoharan, Nature \textbf{483}, 306 (2012).
\bibitem{PRNAR16}
S. Paavilainen, M. Ropo, J. Nieminen, J. Akola,  E. R{\"a}s{\"a}nen, Nano Letters. \textbf{16}, 3519-3523 (2016) 
\bibitem{KGSSR18}
R. Klages, S. Gil-Gallegos, J. Solanp{\"a}{\"a}, M. Sarvilahti, E. R{\"a}s{\"a}nen, {\tt arXiv: 1811.06976}  (2018).
\bibitem{MaZw83}
J. Machta and R. Zwanzig, Rev. Lett. \textbf{50}, 1959-1962 (1983) 
\bibitem{solanpaa2016bill2d}
J. Solanp{\"a}{\"a}, P. Luukko, E. R{\"a}s{\"a}nen, Phys. Commun. \textbf{199}, 133-138 (2016)
\bibitem{RKZ99}
V.~Rom-Kedar, G.~Zaslavsky, Chaos \textbf{9} 697-705, (1999)
\bibitem{thesisGil}
S. Gil-Gallegos, Ph.D. thesis, Queen Mary University of London, 2018
\bibitem{HaCo18}
M. Harsoula, G. Contopoulos, Phys. Rev. E {\bf 97}, 022215 (2018)
\bibitem{CGLS14b}
G. Cristadoro, T. Gilbert, M. Lenci, and D.~P. Sanders,
Europhys. Lett. {\bf 108}, 50002 (2014)

\end{thebibliography}

%

\end{document}